\documentstyle[psfig,12pt]{article}

\newcommand{\eps}{\varepsilon}

\newcommand{\p}{\partial}
\newcommand{\myRe}{\mbox{Re}}
\newcommand{\myIm}{\mbox{Im}}

\begin{document}

\title{\Large
Stability of periodic solutions in series arrays 
of Josephson junctions with internal capacitance}

\author{
Shinya Watanabe\\
\normalsize \it
Niels Bohr Institute (CATS) \\
\normalsize \it
Blegdamsvej 17, Copenhagen \\
\normalsize \it
DK-2100, Denmark \\
\normalsize \it
shinya@alf.nbi.dk \\
\and
James W. Swift\\
\normalsize \it 
Department of Mathematics \\
\normalsize \it
Northern Arizona University \\
\normalsize \it
\normalsize \it Flagstaff, AZ 86011-5717 \\
\normalsize \it
Jim.Swift@nau.edu
}

\date{\normalsize resubmitted to the Journal of Nonlinear Science\\
October 30, 1996}

\maketitle

\begin{abstract}
A mystery surrounds the stability properties of the splay-phase
periodic solutions to a series array 
of $N$ Josephson junction oscillators.
Contrary to what one would expect from dynamical systems theory,
the splay state {\em appears} to be neutrally stable 
for a wide range of system parameters.
It has been explained why the splay state must be 
neutrally stable when the Stewart-McCumber parameter $\beta$
(a measure of the junction internal capacitance) is zero.
In this paper we complete the explanation of the apparent neutral stability;
we show that the splay state is typically hyperbolic --- either asymptotically
stable or unstable --- when $\beta > 0$.
We conclude that there is only a single unit Floquet multiplier,
based on accurate and systematic computations 
of the Floquet multipliers for $\beta$
ranging from $0$ to $10$.
However, $N-2$ multipliers are {\it extremely} close to 1 for $\beta$ 
larger than about 1.
In addition, two more Floquet multipliers approach 1 as $\beta$ becomes large.
We visualize the global dynamics responsible 
for these nearly degenerate multipliers,
and then estimate them accurately by a multiple time-scale analysis.
For $N=4$ junctions the analysis also predicts
that the system converges toward
either the in-phase state, the splay state, 
or two clusters of two oscillators, depending on the parameters.
\end{abstract}

\begin{center}
Abbreviated Title:
Stability of periodic solutions in Josephson arrays.
\end{center}
\newpage

\baselineskip = 7mm

\section{Introduction}
\label{sec:intro}

Arrays of Josephson junctions have been studied intensely 
by mathematicians and physicists because
of their interesting properties as dynamical systems.
A single Josephson junction oscillator is analogous to the familiar physical
pendulum, and the coupling of an arbitrary number of oscillators allows us to
study the dynamics of the extended systems and
the connection between ordinary differential equations and partial
differential equations.
On the other hand, there is also a large
practical interest in Josephson junctions,
which stems from their extremely high frequencies and sensitivity to
the magnetic field \cite{JJA88,ASC94}.
The typical oscillation amplitude of a single junction
is so small that many junctions must be coupled
together in order to get a macroscopic 
voltage oscillation \cite{Jai&al84,Had&al88,Ben&Bur91}.

One-dimensional arrays of Josephson junctions may be classified as
either series or parallel.
Parallel arrays are well approximated by 
a nearest-neighbor diffusive coupling.
This leads to discrete sine-Gordon models,
which support ``solitons'' propagating
across the junctions (see \cite{Wat&al95} and references therein).
In series arrays, the junctions are coupled 
in an all-to-all fashion.
For this ``global'' coupling,
the spatial coordinate is irrelevant, unlike in parallel arrays,
and particularly good analytical progress has been made.

\begin{figure}[tb]
\centerline{\hbox{
\psfig{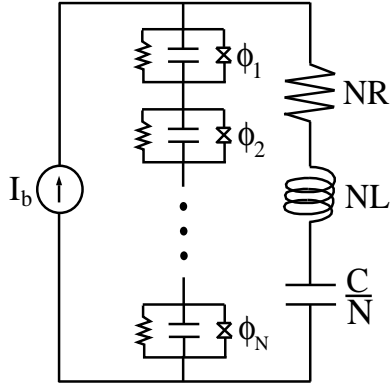}
}}
\caption[circuit]{
\label{circuit}
Circuit diagram for a series array
of $N$ identical Josephson junctions
subject to a constant bias current $I_b$,
and shunted by a series $LRC$ load.
}
\end{figure}

In this paper,
we consider series arrays of $N$ identical Josephson junctions
shunted by a series $LRC$ load (Fig.~\ref{circuit}).
After standard non-dimensionalization,
the governing equations reduce to~\cite{Had&al88}:
\begin{equation}
  \left\{ \begin{array}{l}
    \begin{displaystyle}
      \beta \ddot{\phi_j} + \dot{\phi_j} + \sin \phi_j
        + \dot{Q} = I_b ~, ~~ j = 1, \dots, N ,
    \end{displaystyle} \\
    \begin{displaystyle}
      L \ddot{Q} + R \dot{Q} + \frac{Q}{C} =
                \frac{1}{N} \sum_{k=1}^N \dot{\phi_k} .
    \end{displaystyle}
  \end{array} \right.
\label{goveqn}
\end{equation}
Here, $\phi_j$ is the phase difference across the $j$-th junction,
$I_b$ is the DC bias current, and 
$N L$, $N R$, $C/N$ are the
load inductance, resistance, and capacitance, respectively.
(Thus $L$ is the inductance per junction, etc. 
This scaling, introduced in \cite{Had&al88},
is convenient for comparing different numbers of junctions.)
As mentioned in the previous paragraph, the junctions are globally coupled
even though they are arranged in series.
Mathematically, the equations have permutation symmetry
with respect to the indices $j$. 
The only nonlinearity is the $\sin \phi_j$ term.  In particular, the second
equation is linear.
We emphasize that the bias current $I_b$ is constant in time.  
Josephson junction arrays are often driven by
such a DC forcing since it can still generate useful AC responses.
In addition the equations
become more difficult to analyze with an AC bias current.
(The US voltage standard uses a series array of Josephson
junctions with an AC bias current:  The average voltage across the junctions is
proportional to the frequency of the bias current due to the Josephson
relation~\cite{Kau&al87}.)

The dimensionless Stewart-McCumber parameter $\beta$
is a measure of the junction internal capacitance.  
Junctions can be made with a wide range of $\beta$, ranging from $10^{-6}$ 
to $10^7$~\cite{JJA88,ASC94,vdZ&al94}.
Much progress has been made in analyzing the system (\ref{goveqn}) 
in the limit $\beta = 0$, 
where the phase space of each oscillator is a circle
\cite{Jai&al84,Tsa&al91,Swi&al92,Tsa&Sch92,Str&Mir93,Wat&Str94}. 
In this paper we study the more difficult case when $\beta > 0$, where the 
shape of the oscillation in the $(\phi_j,\dot{\phi_j})$ plane is affected by
the coupling.

A mechanical analog of a {\em single} Josephson junction 
(Eq.(\ref{goveqn}) without $\dot{Q}$) is a damped pendulum
of mass $\beta$, with applied torque $I_b$.  
The coupling in Eq.(\ref{goveqn}) arises in the mechanical analog if
the pendula slide with friction around a common axle,
and the axle is itself a torsional harmonic oscillator
($Q$ is the angle of the axle,
$L$ is proportional to the moment of inertia, etc.).

In this paper we consider only periodic oscillations that are overturning,
i.e.
$\phi_j(t+T) = \phi_j(t) + 2 \pi$ for all $t$ and all $j$
where $T>0$ is the period.
The most important periodic solutions of system (\ref{goveqn}) are
the {\em in-phase} and the {\em splay-phase} states.
The junctions oscillate coherently in the in-phase state:
$\phi_j (t) = \phi_1(t)$ for all $j$.
The splay state is also periodic, but
the individual waveforms are staggered equally in time:
$\phi_j (t) = \phi_1(t +  (j-1)T/N)$ for all $j$.
(This is one splay state;
the indices $j$ can be exchanged arbitrarily
due to the permutation symmetry of Eq.(\ref{goveqn}).)
The in-phase state, due to its obvious coherence,
has been considered promising
as a high-frequency oscillator~\cite{Lik86,Had&al88,Mey&Kre90}.
On the other hand the splay state emits little rf power output,
but other applications are sought,
e.g.\ as a voltage amplifier~\cite{Ter&Bea94}.
Since the existence of a single splay state automatically implies 
the existence of $(N-1)!$ splay states by the permutation symmetry,
the possibility of using these states as multiple-memory device has
also been considered~\cite{Ots91,Sch&Tsa94}.

The linear stability of periodic
solutions is quantified
in terms of Floquet multipliers.
A periodic solution of an ordinary differential equation (ODE) in $n$
dimensions has $n$ Floquet multipliers.  One of these multipliers 
is always $1$, corresponding to the trivial perturbation along the solution, 
and the other $n-1$
multipliers are the eigenvalues of the fixed point of the
Poincar\'{e} map.  (The periodic solution of the ODE is a fixed point of
the Poincar\'{e} map.)  A Floquet multiplier inside the unit circle in the
complex plane indicates a decaying perturbation, while those outside the
unit circle are ``unstable.''

The linear stability of the in-phase state was numerically 
studied \cite{Had&Bea87,Had&al88}
for various combination of the loads, the bias current $I_b$, and $\beta$.
The results were explained analytically in the limit
$\beta=0$ \cite{Jai&al84,Lik86}, 
then for a finite $\beta>0$ \cite{Mey&Kre90,Che&Sch95}
using perturbation methods.
The multipliers of the splay state were also studied intensively,
as we review in the following.
Roughly stated, 
the linear stability properties of the two solutions appear
complementary;
for example, when the in-phase state is stable,
the splay state is usually unstable
(although bistable situations or chaotic attractors may sometimes arise).

However, many numerical studies made a puzzling observation:
often, the splay state is not asymptotically stable
when the in-phase state is unstable;
rather, it is only {\it neutrally} stable
(i.e.\  Liapunov stable but not asymptotically stable).
For an $R$-load with $\beta = 0$, Tsang {\em et al.}~\cite{Tsa&al91}
found evidence that there are $N$ Floquet
multiplers on the unit circle (one complex conjugate pair 
and $N-2$ unit multipliers).
Tsang and Schwartz found that there are
$N-2$ unit multipliers
for the $\beta = 0$ junctions coupled with an
$LC$-load \cite{Tsa&Sch92}.
Nichols and Wiesenfeld \cite{Nic&Wie92} did an excellent study of
five different cases: two with $\beta = 0$ and three with $\beta > 0$. 
They always found, to within their numerical accuracy,
$N-2$ unit multipliers when $\beta = 0$.
For $\beta > 0$, they found that the splay state seems to be 
neutrally stable in $N-2$ directions
for an $LC$ or an $LRC$-load, but asymptotically stable for a $C$-load. 
This difference is surprising because the $C$-load is 
a limiting case of the $LC$-load.
Nichols and Wiesenfeld were very careful when stating their conclusions.
They warned that numerical evidence cannot {\em prove} neutral stability
because one cannot distinguish between a unit multiplier and one that is
extremely close to 1.

All previous analytical studies of this peculiar phenomenon have assumed
$\beta = 0$ for simplicity.  The neutral stability when $\beta = 0$
can be understood in the weak coupling limit \cite{Swi&al92,Wie&Swi95}
or in the limit $N \rightarrow \infty$ \cite{Str&Mir93}.
A rigorous explanation was obtained in
\cite{Wat&Str94}, where it was shown that
the phase space of the system (\ref{goveqn})
has a simple foliated structure when $\beta=0$.  
This structure gives rise to $N-3$ constraints 
(constants of motion) to the system, 
and hence imposes the splay solution to have $N-2$ unit multipliers.
This {\it global} result not only explained the peculiar linear stability
of the splay state,
but also made it possible to analyze the asymptotic behavior
from an arbitrary initial condition.
(Because initial conditions of the junctions cannot be specified,
it is important that a desired periodic solution is globally stable, 
not just linearly stable, for any applications \cite{Bra&Wie94}.)
In the weak coupling limit \cite{Wat&Str94}
it was shown that the possible asymptotic states are
either the in-phase state or one of the ``incoherent states''
which include the splay state as a special case.

The present paper studies, numerically and then analytically,
what happens to the phase space structure when $\beta$ is positive,
and connects the global picture to the local stability properties
of the in-phase and the splay periodic solutions.
Since Nichols and Wiesenfeld found apparent neutral stability in two
of the three cases when $\beta > 0$,
an early conjecture was that the simple structure of the $\beta = 0$
equations might persist for $\beta > 0$,
and one might hope to {\em prove} 
that the splay state is still neutrally stable.

But as we show numerically in Section 2, 
the splay solution is
{\em not} neutrally stable when $\beta > 0$ with an $LC$-load. 
Unlike Nichols and Wiesenfeld \cite{Nic&Wie92},
we choose $L=1$ and $C=1/4$
since this is ``closer'' to a $C$-load 
($1/C$ is large compared to $L$).
We use $N=4$ junctions since this is the smallest $N$ 
that shows the degenerate structure of the $\beta=0$ equations.
We visualize the global dynamics by using the time delay technique,
and find that the splay solution is asymptotically stable
for some set of $\beta$ and $I_b$.
This leads us to suspect that the neutral stability observed 
when $\beta > 0$
in other regions of the $(L, C)$ parameter space is 
also only approximate.
Furthermore, an $LRC$-load is qualitatively similar to an $LC$-load;
the system is dissipative even with an $LC$-load due to
the internal resistance of the junctions.

The visualization of the flow in Sec.~2 also reveals that
there are several different time scales in the dynamics
when the splay state is stable.
A trajectory initially converges to a manifold of ``incoherent states'',
then slowly drift toward the splay state.
This process can be quantified by monitoring
the ``moments'' $R_m$ of the phases
(defined in Sec.~2.2).
The first moment $R_1$ converges to 0 rapidly, 
then the second moment $R_2$ slowly drift to 0.
By choosing another set of $\beta$ and $I_b$,
we can change the direction of the drift,
while still forcing the incoherent manifold to be stable.
$R_1$ still converges to 0, but then $R_2$ drifts toward 1.
The final state is a periodic state
with 2 antiphase pairs of oscillators.
For yet another $\beta$ and $I_b$, $R_1$ converges to 1,
in which case the in-phase state is attracting.

For $\beta=0$, the drift of $R_2$ is prohibited 
due to  the rigid foliation structure;
the drift is an important effect of positive $\beta$.
This change in the global dynamics is reflected in 
the Floquet multipliers around the splay state.
Generically, there should be a single unit Floquet multiplier, 
but there are $N-2$ unit multipliers when $\beta = 0$.
For $\beta>0$, $N-3$ of them are perturbed away from the unity,
and the genericity is recovered.
With $N=4$, only one multiplier is perturbed,
which we call the ``critical multiplier.''

To test this qualitative picture more quantitatively,
we present in section 3 systematic
computations of the Floquet multipliers as a function of $\beta$.
We choose $N=4$ and the same load values
as used by Nichols and Wiesenfeld \cite{Nic&Wie92}
for both the $LC$ and $C$-load.
We continue the branch of splay solutions 
between $\beta = 0$ and $\beta = 10$
using AUTO86 \cite{Doe&Ker86}.
This allows us to track the single critical multiplier
into the regime where the splay solution is unstable.
We find that this multiplier approaches 1 as $\beta \rightarrow 0$, as
expected.  However this multipler is
also asymptotic to 1 as $\beta \rightarrow \infty$.  In fact, the multiplier
approaches 1 so quickly that Nichols and Wiesenfeld could not
distinguish it from 1 when $\beta = 1.1$ in the $LC$-load case.
(Our calculations are consistent with theirs.)
At smaller $\beta$, where the critical multiplier is significantly smaller
than 1, 
the splay state is unstable in a different direction and thus
Nichols and Wiesenfeld did not find the state 
with their forward integration.
On the other hand, in the $C$-load case 
it turns out that the splay state is
stable for $\beta$ small, and they observed an asymptotically stable
splay state at $\beta = 0.1$.
For both loads,
we also find that $N$ Floquet multipliers approach 1 
as $\beta \rightarrow \infty$, 
indicating that the junctions become uncoupled
in this limit.

There are two limits in which the oscillators become
uncoupled regardless of the load:
$\beta \rightarrow \infty$, or $I_b \rightarrow \infty$. 
(Furthermore, the coupling is weak when the load impedance becomes large,
for example $R \rightarrow \infty$.)
We follow \cite{Che&Sch95}, and consider the limit $I_b \rightarrow \infty$.
In section 4 we use the perturbation parameter $\eps = 1/I_b \ll 1$
to analytically verify the global picture 
visualized in Sec.~2,
and estimate the Floquet multipliers computed in Sec.~3.
Based on our observation in Sec.~2,
we introduce three different time scales,
and derive the order $\eps^4$ averaged equations
for the phase drift of the oscillators.

In Sec.~4.3 the order $\eps^2$ averaged equation is obtained,
which turns out to have the same form as the averaged equation
when $\beta=0$ \cite{Wie&Swi95}: it predicts that the splay solution has
$N-2$ unit multipliers.
While for $\beta=0$ this approximate result can be made
rigorous \cite{Wat&Str94},
we expect higher order corrections to all but one unit multiplier.
The $O(\eps^2)$ averaged equation controls the 
flow between the in-phase state
and the incoherent manifold.
The first moment $R_1$ is shown to converge toward 0 or 1,
depending on parameters.
The stability of the splay solution to perturbations of $R_1$
corresponds to a single complex conjugate pair of Floquet multipliers.
We estimate this multiplier pair from the second order averaged equation
and find excellent agreement with our numerical results
around the splay state.

In order to resolve the drift along the incoherent manifold,
we expand to higher order and derive the order $\eps^4$ corrections
to averaged equation in Sec.~4.4.
The second moment $R_2$ now converges to 0 or 1.
We compute the critical multiplier for $N=4$,
and again find excellent agreement with our numerical data.

A summary and a discussion are presented in Sec.~5.
Appendices A and B show details of the analysis in Sec.~4.


\section{Numerical integration of the equations}

In this section we numerically integrate the equations (\ref{goveqn})
using a standard ODE integrator.
This has the disadvantage that we cannot find unstable states.  Nonetheless
we will convince the reader that the degeneracy observed when $\beta = 0$ does
not persist when $\beta > 0$, although the degeneracy re-appears {\em
approximately} when $\beta$ is large.

Since Nichols and Wiesenfeld~\cite{Nic&Wie92}
observed an asymptotically stable 
splay state in the $C$-load case, 
we are interested primarily in the $LC$ and $LRC$ loads.  These are
not qualitatively different in the Josephson junction system because of the
internal resistance of the junctions.  Thus, we only consider the $LC$ load.

We shall find that in some parameter regions of the $LC$-load system, there are
asymptotically stable splay solutions, which persist when a small nonzero
resistance is added to the load.  We will also show other types of periodic
solutions that can be attracting.

\subsection{Time-delayed coordinates}

Both the in-phase and the splay states are {\em phase-locked}
states,
and the phase relations among junctions are defined
in terms of time lags rather than mere differences 
in angles $\phi_j$
at each instant $t$
since the waveform $\phi_j (t)$ is nonuniform.
A process to convert a solution $\phi_j (t)$ 
into the phase information is therefore necessary.

One useful algorithm~\cite{Ash&Swi93} is based on time delays
between the ``firings'' of oscillators.
We define each junction to ``fire''
when $\phi_j = \phi^*$ (mod $2 \pi$)
for some fixed $\phi^*$.
(We choose $\phi^* = 0$.)
Let $t_j (p)$ be the time of the $p$-th firing 
of the $j$-th junction.
We monitor firings of all the junctions,
and calculate the phases every time junction $N$ fires.
Assume that $p$ and $q$ satisfy
$t_j (q) \leq t_N (p) < t_j (q+1)$.
Then we define the phase (normalized time lag)
of the $j$-th junction to be:
\begin{equation}
  \theta_j (p) = 
  2 \pi \left( \frac{t_N (p) - t_j (q)}{T_j} \right)
\label{delaycoord}
\end{equation}
where $T_j$ is an approximate period of the junction.
The phase $\theta_j(p)$ is approximately the angle that
oscillator $j$ covers between its most recent firing and the $p$th firing of
oscillator $N$.
Among the candidates for $T_j$,
we choose $T_j = t_j (q+1) - t_j (q)$ in the following.
This has an advantage that $\theta_j$ always falls in the interval
$0 \leq \theta_j < 2 \pi$.
The ``time-delayed coordinate'' $\theta_j$ is an angle coordinate
and we identify $\theta_j = 0$ with $\theta_j = 2 \pi$.

\subsection{Periodic attractors}

We integrate the system (\ref{goveqn}) 
with $N=4$
by an adaptive time-step Runge-Kutta integrator,
and compute the time-delayed coordinates.
We have tested both $LC$ and $C$ loads,
but will focus on one representative case
when $L=1$, $R=0$, $C=1/4$.
An array with this load exhibits a variety of periodic solutions
depending on the parameters $(I_b,\beta)$.
The initial conditions are chosen to illustrate the
transient dynamics.

\begin{figure}[tb]
\centerline{\hbox{
\psfig{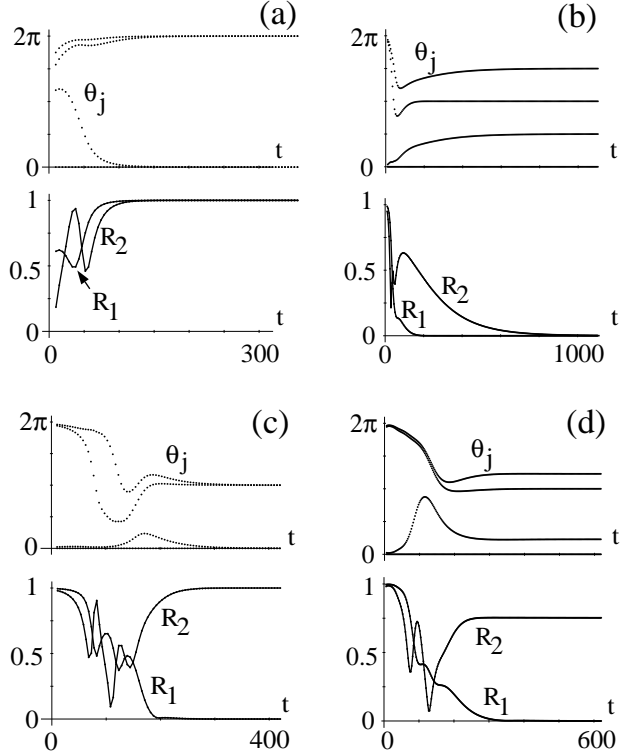}
}}
\caption[delayfig]{
\label{delayfig}
Time-delayed coordinates $\theta_j$ (upper graphs) and
the amplitude of the first and the second moments
$R_1$ and $R_2$ (lower graphs)
as functions of $t$, for $N=4$ Josephson junctions.
Cases (a--c) uses $L=1$, $R=0$, $C=1/4$, and $I_b=1.5$.
Only the McCumber parameter $\beta$ is changed:
(a) 0.7, (b) 0.15, (c) 0.3.
Case (d) uses the same load but with $I_b=2.5$ and $\beta=1$.
A trajectory converges to (a) the in-phase state,
(b) the splay state, (c) the 2+2 incoherent state,
and to (d) a (generic) incoherent state.
}
\end{figure}

Four cases are shown in Fig.~\ref{delayfig}.
For each case, two graphs are shown;
an upper graph shows the time-delayed coordinates $\theta_j$.
Since $\theta_N \equiv 0$ by definition,
only three curves can be seen.
Each lower graph shows the amplitudes
$R_1$ and $R_2$ of the
the first and the second (complex) moments $z_m$
of $\theta_j$'s:
\begin{equation}
 z_m =
  \frac{1}{N} \sum_{k=1}^N e^{i m \theta_k }~,~~
  R_m = | z_m | ~.
\label{defordpar}
\end{equation}
The amplitude of the first moment, $R_1$,
was originally introduced by Kuramoto as an ``order parameter''
to measure the phase coherence 
of a population of phase oscillators \cite{Kur84}.
Indeed, $R_1=1$ if and only if the oscillators are in-phase.
When $R_1 = 0$, we say that the oscillators are ``incoherent'' (this is our
{\em definition} of an incoherent state).
The splay state is characterized by $R_m = 1$ if $m$ is a multiple of $N$,
and $R_m = 0$ otherwise.
The horizontal $t$-axis
for the figures
denotes the (discrete) firing time $t_N (p)$.
The first several firings are ignored
since the junctions rapidly change their phases.
During this transient,
the time-delayed coordinates
are not meaningful.

For Figs.~\ref{delayfig}(a) to (c),
the bias current is fixed at $I_b=1.5$,
and only the McCumber parameter $\beta$ is changed.
Fig.~\ref{delayfig}(a),
which shows a stable in-phase attractor, uses $\beta=0.7$.
Since $\theta_j=0$ is identified with $\theta_j=2\pi$,
we see that four junctions have become synchronized after $t \approx 100$.
This is also clear from the moments:
both $R_1$ and $R_2$ converge to 1.

In Fig.~\ref{delayfig}(b), $\beta=0.15$ is used.
The phase differences converge to $2 \pi/N$,
thus the attractor is the splay state.
The way the trajectory approaches the attractor is somewhat interesting.
Note that $R_1$ vanishes at $t \approx 200$, after which
the state is incoherent.
If the angles $\theta_j$ are plotted on the unit circle,
they are scattered so that their center of mass lies
at the origin.
For $N=4$, an incoherent state has two ``antiphase'' pairs:
The oscillators can be ordered so that
\begin{equation}
\label{incoherent}
(\theta_1, \theta_2, \theta_3, \theta_4) = (\theta_1, \pi, \theta_1+\pi, 0)
\end{equation}
with $0 \leq \theta_1 \leq \pi$.
After $R_1$ vanishes,
only the angle $\theta_1$ (and hence $\theta_3$)
in equation~(\ref{incoherent}) changes,
and the incoherent state is characterized by
$R_2 = | \cos \theta_1 |$.
Note the {\em slow drift} of $R_2 (t)$ toward $0$,
indicating the splay state with $\theta_1 = \pi/2$.

Fig.~\ref{delayfig}(c) uses an intermediate value $\beta=0.3$.
At $t \approx 200$, $R_1$ vanishes and the phases are 
again incoherent.
A slow drift of $R_2$ follows, as in (b) occurs, but in the opposite direction.
At $t \approx 300$,
$R_2$ converges to unity,
indicating that $\theta_1$ approaches $0$ or $\pi$.
The final state is two clusters located 180 degrees apart.
We call this state the ``2+2 incoherent'' periodic solution.

With $I_b = 1.5$ as in Figs.~\ref{delayfig}(a--c),
we did a parameter sweep of $\Delta \beta = 0.01$, 
and found that
the splay state is stable for $\beta \leq 0.21$,
the 2+2 incoherent state is stable for $0.19 \leq \beta \leq 0.61$,
and the in-phase state is stable for $\beta \geq 0.40$.
This bistability between the splay and 2+2 incoherent states implies
that there exists
an unstable generic incoherent state
when $0.19 \le \beta \le 0.21$.
By ``generic,'' we mean that $0 < \theta_1 < \pi/2$ in
equation~(\ref{incoherent}).

Fig.~\ref{delayfig}(d) uses the same load,
but both $I_b = 2.5$ and $\beta=1$ are larger than the previous values.
We see that $R_1$ vanishes at $t \approx 300$ again
while $R_2$ stays constant for $t > 250$.
The constant value is neither 0 nor 1
so that the final state appears to be a generic
incoherent state with $0 < \theta_1 < \pi/2$.
Different initial conditions converge to incoherent states with different
$R_2$ (and thus different $\theta_1$.)
It {\em appears} that there is an incoherent periodic state for each value or
$R_2$ (or $\theta_1$), just as there is in the equations with $\beta = 0$.
However, we believe that the states are {\em not} exactly periodic but are
drifting very slowly towards the splay state.
In Sections 3 and 4 we will give evidence for this slow drift.

When $I_b$ or $\beta$ is larger, for example $I_b = 2.5$, $\beta = 4$
or $I_b = 4$, $\beta = 1$,
there appears to be a periodic state with {\em any}
values of the angles $\theta_i$.
During the transient dynamics the delay coordinates are not meaningful,
but after this the angles $\theta_j$ stay constant at
values determined by the initial conditions.
In other words, the oscillators are effectively
decoupled, and they maintain any phase relationship.  Mathematically speaking,
there appears to be an attracting 4-torus foliated by periodic orbits.
As with the case described in figure~\ref{delayfig}(d), 
however, we expect that
the dynamics are actually ``generic'',
but the drift of both $R_1$ and $R_2$
occurs on a time scale too long to observe.

All of the oscillators have the same waveform in the
in-phase, splay, and 2+2 incoherent solutions.
The existence of such solutions has been proved in
\cite{Aro&Hua95} for a wide range of loads, including ours.
In a generic incoherent state implied above, however,
one antiphase pair has a different waveform from the other pair
(see section 2.4)
and there is no existence proof for this type of periodic solutions.

\subsection{Three-dimensional projection}

\begin{figure}[tb]
\centerline{\hbox{
\psfig{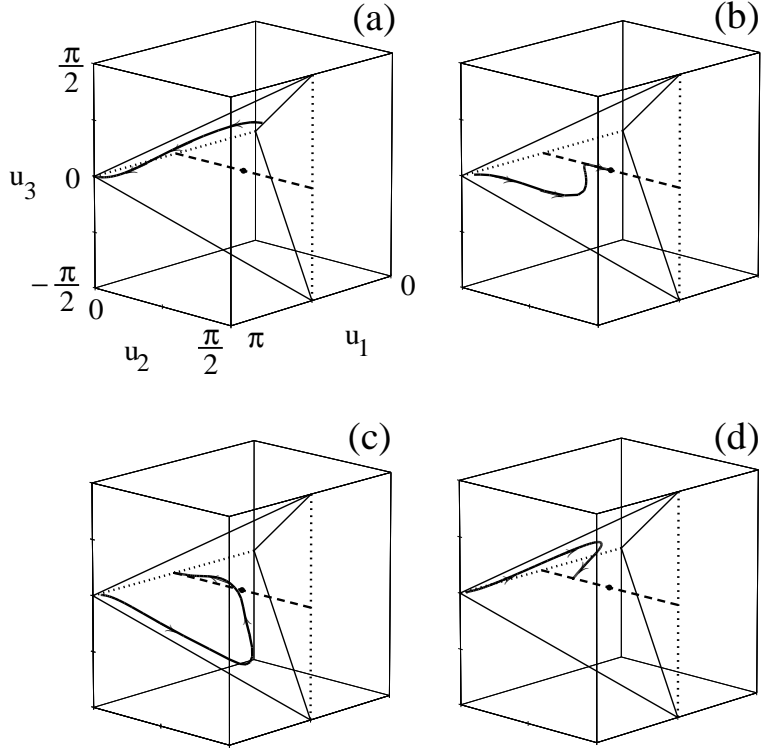}
}}
\caption[3dfig]{
\label{3dfig}
Projection of trajectories onto a three-dimensional space
defined by a linear transformation of the time-delay coordinates:
$u_{0,1} = \theta_1 + \theta_2 \pm \theta_3 \pm \theta_4$
and
$u_{2,3} = \pm \theta_1 \mp \theta_2 + \theta_3 - \theta_4$.
Only the phase {\em differences} $u_1$, $u_2$, $u_3$ are shown.
The cases (a--d) correspond to the cases in Fig.~\ref{delayfig},
respectively.
The vertices of the tetrahedron correspond to the in-phase state
while the bar inside the region is the set of incoherent states.
The two endpoints of the bar are both 2+2 incoherent states,
while the dot in the middle of the bar is the splay state.
}
\end{figure}

The dynamics become more apparent by
using a linear transformation of the time-delayed 
coordinates~\cite{Ash&Swi92,Tsa&Sch92,Wat&Str94}.
Since we are only interested in the relative phase differences,
we change the coordinates by, for example,
$u_{0,1} = \theta_1 + \theta_2 \pm \theta_3 \pm \theta_4$
and 
$u_{2,3} = \pm \theta_1 \mp \theta_2 + \theta_3 - \theta_4$,
then only show $u_1$, $u_2$, and $u_3$.
Figs.~\ref{3dfig}(a--d) show the dynamics 
of Figs.~\ref{delayfig}(a--d), respectively, in these coordinates.

The trajectories run inside 
a tetrahedron in this representation.
The faces of the tetrahedron correspond to when two $\theta_j$'s
are identical (a ``2+1+1'' state).
Each edge of the tetrahedron corresponds to one of two types of oscillation:
``2+2'' (shown as dotted lines) or ``3+1'' (as solid lines).
Since no change in ordering of junctions occurs
in Fig.~\ref{delayfig} 
after a short transient,
the phases maintain the relationship, say,
$\theta_4  = 0 \leq \theta_3 \leq \theta_2 \leq \theta_1 \leq 2 \pi$.
Thus, a trajectory does not escape from the tetrahedron.
(Such an escape is prohibited
{\em if} there is an attracting invariant $4$-torus.  However, with
strong coupling the oscillators can shuffle their order
even after they have converged to the attractor.)
The four vertices of the tetrahedron correspond to the in-phase state.
(There are four ways to approach this state
while maintaining the ordering.)
The {\em incoherent bar} inside the tetrahedron 
is the set of incoherent states
for which $R_1$ vanishes.
(The incoherent bar, indicated by a dashed line, generalizes to an
$(N-3)$-dimensional incoherent manifold when there are $N$ oscillators.)
The two end points of the bar are
2+2 incoherent states,
while the dot at the midpoint is the splay state.
The trajectories shown are actually discrete sets of points
at $\{ t_N (p) \}$,
but the points are connected in the figures for clarity.

In Fig.~\ref{3dfig}(a)
the trajectory starts at a point inside the tetrahedron,
but converges to the in-phase state at a vertex.
In Fig.~\ref{3dfig}(b), on the other hand, 
the initial condition is near the in-phase state
but the trajectory converges to the splay state.
This process consists of two stages;
the first fairly rapid fall onto the incoherent bar
is followed by a slow drift along the bar.

In Fig.~\ref{3dfig}(c)
the initial condition is also near the in-phase state,
and the trajectory falls onto the incoherent bar.
However, this time, it converges to the 2+2 incoherent state
at the end of the bar.
In Fig.~\ref{3dfig}(d)
the trajectory does not appear to move along the incoherent bar
once it hits the bar.
The calculation is halted at $t \approx 600$.
However, we expect that there is 
a very slow drift along the bar as described earlier.

The drift along the incoherent bar
implies that the degenerate phase space structure of the $\beta=0$ limit
is broken for $\beta>0$.
In this limit,
the tetrahedron in Fig.~\ref{3dfig} 
is foliated by a family of invariant subspaces~\cite{Wat&Str94}.
If such degeneracy were to persist when $\beta > 0$,
then the structure would prohibit a trajectory from
moving from one subspace to another,
thus the drift would not occur.
As a result of this drift,
the splay state, the 2+2 incoherent state, 
or more generic incoherent states
can become asymptotically stable.

Is the incoherent manifold dynamically invariant?
For $N=3$ the incoherent manifold is precisely the splay state,
so the answer is yes.
Appendix B gives a formal proof that the incoherent manifold is
dynamically invariant for $N=4$ and weak coupling.  The proof is not rigorous
because it relies on the phase-shift symmetry induced by averaging.
In Appendix B we also show that 
the incoherent manifold, defined by $R_1 = 0$, is
not invariant for $N > 4$.
However, when the coupling is weak there is an invariant manifold,
close to the set $R_1 = 0$, on which there is a slow drift.
Thus, the concept of an incoherent manifold is still useful.

\subsection{Waveforms}

The time-delayed coordinates $\theta_j$ are useful
to investigate the dynamics,
but do not show direct information on $\phi_j (t)$.
Therefore, we investigate the waveforms
and symmetry properties of the periodic solutions
in this section.

\begin{figure}[tb]
\centerline{\hbox{
\psfig{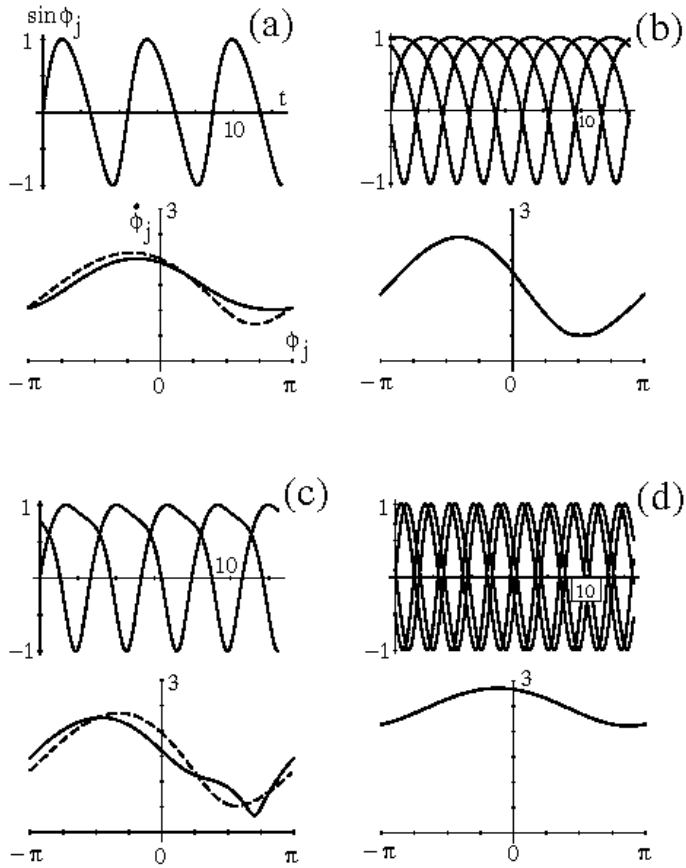}
}}
\caption[waveform]{
\label{waveform}
The supercurrent $\sin \phi_j (t)$ (upper graphs) and
the phase portrait $\dot{\phi_j}$ versus $\phi_j$ (lower graphs)
of the final state achieved in Fig.~\ref{delayfig}, respectively.
The dashed curve in the phase portrait shows
the trajectory of an uncoupled junction
with the same $I_b$ and $\beta$ values.
}
\end{figure}
 
Figures \ref{waveform}(a--d) show the waveforms
for (a) the in-phase state, (b) the splay state,
(c) the 2+2 incoherent state, 
and (d) the generic incoherent state, respectively.
(The case (d) is close enough to being periodic.)
Each upper graph displays $\sin \phi_j$ 
(the supercurrent through the $j$-th junction)
as a function of $t$,
whereas each lower graph shows the same solution
in terms of the phase portrait $(\phi_j,\dot{\phi_j})$
for all $j$.
The dashed curve in the lower graph
denotes the trajectory of an uncoupled (single) junction
with the same parameters $(I_b,\beta)$.

The trajectories for the in-phase state 
and the 2+2 incoherent state
are significantly altered from the trajectory without coupling.
In contrast, the trajectories for the splay state
and the generic incoherent state are almost unchanged in shape.
This is partly due to the parameter values;
note that figures (b) and (d) have the largest $\beta$ values.
Furthermore, when the state is incoherent the coupling terms
in equation (\ref{goveqn}), $\sum \dot{\phi_k}$, tend to cancel each other.

In all four cases, the four trajectories in the phase
portraits follow the same curve.
This must be the case for all but the generic incoherent state in figure (d).
Let us explain:  We have already given the general form of the in-phase,
splay, and 2+2 incoherent solutions.  Each of these states has a single
waveform, with different phase shifts between the oscillators.
For any incoherent state with $N=4$, the oscillators can be numbered so that
$\phi_3(t) = \phi_1(t + T/2)$, and $\phi_4(t) = \phi_2(t+T/2)$, where $T$ is
the period.
In figure (d),
we observe that $\phi_2(t) = \phi_1(t+c)$ for some phase shift $c$, but in
general these two waveforms can be different.
We have in fact observed two distinct waveforms for a generic incoherent state
in a different parameter regime.

\begin{figure}[tb]
\centerline{\hbox{
\psfig{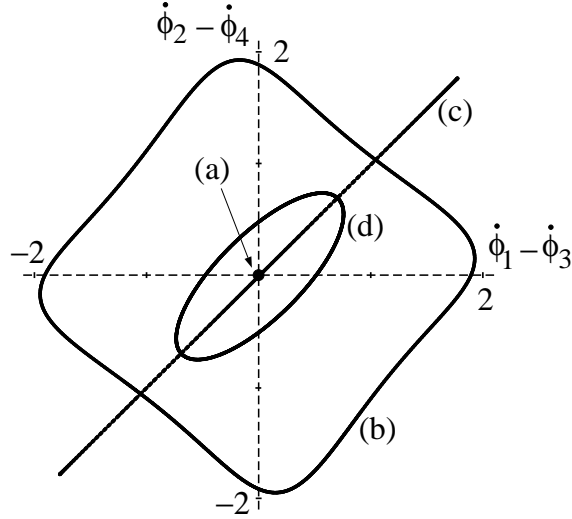}
}}
\caption[lissajous]{
\label{lissajous}
Lissajous figures obtained by plotting the difference of the voltages across
junctions 2 and 4 versus that of junctions 1 and 3.  Four figures are
superimposed, corresponding to the periodic states in
Fig.~\ref{delayfig}.
The in-phase state (a) appears as a point at the origin.
}
\end{figure}

Finally, we show a way to distinguish between the different types of
oscillations which does not require the time-delay technique.
In Fig.~\ref{lissajous} we plot
$\dot{\phi_2}-\dot{\phi_4}$ vs.
$\dot{\phi_1}-\dot{\phi_3}$.
Since $\dot{\phi_j}$ is proportional to the voltage
across the Josephson junction,
we call this the Lissajous figure 
created by the voltage signals.

All $\phi_j (t)$ are synchronized 
for the in-phase solution,
thus the figure for case (a) appears as
the point at the origin.
The pattern of the other states depends on the order of the oscillators: we
choose initial conditions or permute the oscillators so that 
$\phi_1 \le \phi_2 \le \phi_3 \le \phi_4 \le \phi_1 + 2 \pi$.
The splay state (b) has a 4-fold symmetry
while a generic incoherent state (d) has only
2-fold rotational symmetry about the origin.
The 2+2 incoherent state appears as a diagonal line
with a slope $+1$
(The slope can also be $-1$, depending on initial conditions.)
The symmetries of the Lissajous figures 
are easily explained from the waveforms.
For example, phase shifting the splay state by a quarter period 
rotates the Lissajous figure by 90 degrees.

The Lissajous figures can be extended to three
dimensions using the ``$z$" coordinate 
$\dot{\phi_1} - \dot{\phi_2} + \dot{\phi_3} - \dot{\phi_4}$.
The resulting figures are analogous to Fig.~\ref{3dfig},
but they can be made without computing the time-delay coordinates.
The $z$ coordinate of the Lissajous figure is needed to distinguish
between an incoherent oscillation (on the incoherent bar in Fig.~\ref{3dfig})
and a general oscillation (in the interior of the tetrahedron).
Both oscillations look like ellipses in the
$\dot{\phi_2}-\dot{\phi_4}$ vs. $\dot{\phi_1}-\dot{\phi_3}$ plane.
However, incoherent oscillations have the extra symmetry that
$\dot{\phi_1} - \dot{\phi_2} + \dot{\phi_3} - \dot{\phi_4}$
is unchanged after {\em half} a period.


\section{Floquet multiplier computation}
\label{sec:floquet}

In this section we report on calculations of the Floquet multipliers of the 
splay-phase state. The Floquet multipliers provide 
local information which is 
complementary to the images of the global phase space 
shown in the previous section.
The degenerate dynamics of the equations when $\beta = 0$ is reflected by the 
fact that the splay state has $N-2$ unit multipliers when $\beta = 0$.  
Conversely, if the dynamics is nondegenerate,
then there must be exactly one 
unit Floquet multiplier.

We calculate the multipliers of the splay state as a function of $\beta$
for two types of the shunt loads,
an $LC$ load and a purely capacitive load.
We find that there is a single unit Floquet multipler when $\beta >0$, 
indicating that the system is nondegenerate.  
However, when $\beta$ is larger than
about 1, there are $N-2$ multipliers {\em very close} to 1.
This is why we observed an apparently neutrally stable incoherent state in 
Fig.~\ref{delayfig}(d):
the Floquet multiplier which measures the speed of the drift in 
the incoherent manifold is extremely close to 1.

Furthermore, a complex conjugate pair of multipliers approaches 1 as 
$\beta \rightarrow \infty$.  
Thus, in the large $\beta$ limit there are $N$ unit
Floquet multiplers, corresponding to the oscillators 
being effectively uncoupled, 
which we observed in the numerical integrations.

In order to calculate the Floquet multipliers, we first find a stable splay 
state for one set of parameters,
then use a continuation method to compute the
branch of periodic solutions as $\beta$ is varied,
holding the other parameters fixed.
Solutions can be obtained 
regardless of stability,
which clarifies the dependence of the multipliers on $\beta$.

The starting solution is found
by choosing the same parameters
as two of the cases studied by 
Nichols and Wiesenfeld~\cite{Nic&Wie92},
specifically, their cases II and III:
\[
  \begin{array}{lllllll}
    \mbox{[NWII]} & N=4 & L=0.75 & R=0 & C=20 & I_b=2.5 & \beta=1.1 \\
    \mbox{[NWIII]} & N=4 & L=0 & R=0 & C=1 & I_b=1.5 & \beta=0.1
  \end{array}
\]
(We have scaled their actual parameter values
to suit our definition.)

In both cases,
we have chosen an initial condition 
close to the splay state.
Since the waveform is nonuniform,
$\phi_j (0) = 2 \pi j/N$ is not a good starting point.
Instead, we have computed the periodic solution $\phi^* (t)$
of the uncoupled equation
\begin{equation}
  \beta \ddot{\phi_j} + \dot{\phi_j} + \sin \phi_j = I_b
  \label{uncoupled}
\end{equation}
for the given $(I_b,\beta)$,
then assigned the initial condition as
$\phi_j (0) = \phi^* (T j/N)$
where $T$ is the period.
The choice is especially useful for the $LC$-load (case II),
and it enables us to obtain the splay state
(or, at worst, an incoherent state extremely close
to the splay state) sufficiently fast.
We do not need an extrapolation technique as used in \cite{Nic&Wie92}.

Next, we sweep $\beta$ up to $10$ 
using the continuation and bifurcation 
package AUTO86 \cite{Doe&Ker86}.
Then, $\beta$ is decreased from $10$ down to almost $0$.
AUTO86 computes the Floquet multipliers, but not accurately enough for our 
purposes.
To improve the accuracy, 
we output a point on the periodic solution at each value of $\beta$, 
and integrate the Jacobian matrix for one full period.
The eigenvalues of the resulting matrix are the Floquet multipliers, which we 
compute using Mathematica.
There must always be a unit multiplier
corresponding to perturbation 
in the direction of the periodic orbit.
This multiplier is computed to 8--12 significant figures.
(AUTO86, which needs Floquet multipliers
only for detecting bifurcation points,
has calculated the trivial multiplier within 3--7 digits.
An improved version AUTO94 was not available
at the time of our computation.)
Also, the multipliers for the starting value of $\beta$
agree with those obtained in \cite{Nic&Wie92}.

\begin{figure}[tb]
\centerline{\hbox{
\psfig{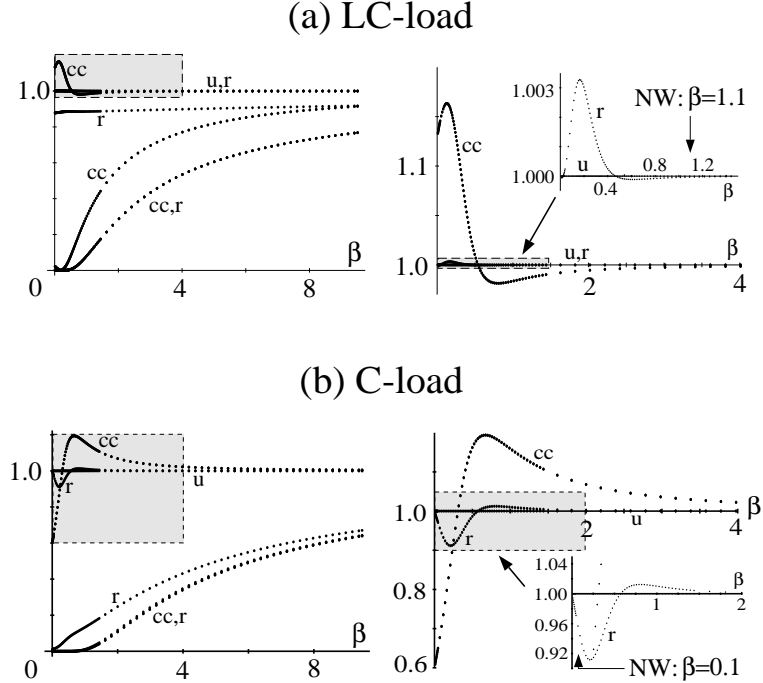}
}}
\caption[fmult]{
\label{fmult}
The magnitude of Floquet multipliers
as a function of $\beta$,
for (a) an $LC$ load (case [NWII])
and for (b) a $C$ load (case [NWIII]).
All multipliers are shown in the left figures
while the shaded regions are enlarged in the right figures.
In order to resolve the real multiplier near unity,
another set of enlargements (of the shaded region
in the right figures) are shown as insets.
There are 10 multipliers for (a) and 8 for (b).
The letters denote ``cc'' (complex conjugate pairs),
``r'' (real), and ``u'' (the unit multiplier).
The arrows labeled ``NW'' point to the $\beta$ values
used in \cite{Nic&Wie92}.
}
\end{figure}
 
Figure \ref{fmult} shows the magnitude 
of the computed multipliers.
The figures (a) and (b) correspond to [NWII] and [NWIII],
respectively.
The left figures show all the multipliers
whereas the right ones are enlarged views 
of the shaded box in the left figure.
To resolve the real multiplier near unity,
the shaded box in the right figure is
further enlarged as the inset.
The number of multipliers is the same as the dimension of
the system, thus there are ten multipliers in (a)
and eight in (b).
They are labeled as ``cc'' (complex conjugate pairs),
``r'' (real), and ``u'' (the unit multiplier).
One sees that the unit multiplier is computed accurately,
and shows no visible deviation from unity even in the inset.

Although these multipliers provide only local information
around the splay state,
we can roughly associate them with the global dynamics
described in Section 2.
For example, five multipliers in (a) and four in (b)
have magnitudes well below unity, 
especially when $\beta$ is small.
They correspond to the rapid initial transient 
when the junctions shuffle and change firing order.
After these directions are contracted,
all the junctions have similar periods.
The convergence speed of this initial stage is instantaneous when $\beta=0$
and becomes slower as $\beta$ becomes large.
One real multiplier in (a) which always stays near $0.9$
seems to correspond to a perturbation in the charge $Q$
on the capacitor.  This multiplier behaves differently from the others.

After the initial transients are over, 
the time-delayed coordinates described in Section 2
become meaningful, and the dynamics involves the phase differences.
The three nontrivial multipliers
of near-unit magnitude, shown in the right panels in Figs.~\ref{fmult}(a,b),
represent this dynamics.
From now on, our interest is focused on these three multipliers.
The incoherent manifold is the key to understanding the dynamics.
Two multipliers (the c.c. pair) correspond to
the dynamics transversal to this manifold.
If they have magnitude greater than unity,
then trajectories depart from the manifold
and the in-phase state is usually stable.
When their magnitude is below unity,
the incoherent manifold is attracting, at least near the splay solution.

The real multiplier near unity,
which we call the ``critical multiplier,''
represents the dynamics along the manifold.
(We describe 4 oscillators.  In general there are $N-3$ critical multipliers.)
When the critical multiplier is less than 1, 
the splay state is attracting 
within the incoherent manifold.
When it is larger than 1, the drift on the incoherent manifold is away 
from the splay state.
In the $\beta = 0$ equations the critical multiplier is exactly unity, 
and there is no drift on the incoherent manifold.

Figure \ref{fmult} shows that the magnitude 
of the ``transverse multipliers'' deviates substantially
from 1 when $\beta$ is small.
The magnitude changes in a nontrivial fashion,
switching stability at finite $\beta$ in either case.
Using an $LC$ load, 
the incoherent manifold is repelling
for small $\beta$, and attracting for larger values.
For a $C$ load, it is reversed.

The critical multiplier is extremely close to (but not equal to)
1 at moderate $\beta$.
Thus it {\em appears}
that degenerate phase space structure
persists for $\beta>0$
\cite{Nic&Wie92}, since the drift on the incoherent manifold is too
slow to observe.
However, we believe that the dynamics is nondegenerate for $\beta > 0$.
At those $\beta$ values where the critical multiplier crosses unity,
we expect a pitchfork bifurcation that creates two generic incoherent
solutions.
Analysis \cite{Wat&Str94} predicts that the real multiplier
must be exactly 1 at $\beta=0$.
Fig.~\ref{fmult} is in agreement.
(Yet another enlargement of the inset of
Fig. 6(a) shows that the critical multiplier
dips below $1$ at very small $\beta$ and then approaches $1$ as
$\beta \rightarrow 0$.)


\section{Analysis}
\label{sec:averaging}

\noindent
In this section we show analytically the weak destruction
of the foliated phase space described in Sec.~2,
and estimate the Floquet multipliers
around the splay state computed in Sec.~3.
The analysis uses the multiple time-scale method,
and is applicable for a general $N$, 
but our primary focus is on the case $N=4$
in order to compare with the previous sections.

We introduce a small parameter $\eps$ and different
time scales in Sec.~4.1,
then study the $O(1)$ equations in Sec.~4.2.
This lowest order system describes the dynamics
when the junction phases are being shuffled and
the time-delayed coordinate is not yet meaningful.
In Sec.~4.3 we go up to the $O(\eps^2)$ expansion, 
and derive an averaged equation,
which describes the drift of the phases.
The order parameter $R_1$ is shown to converge toward either 0 or 1.
The averaged equation can predict the magnitude of the transverse multipliers,
but the critical multiplier is not yet resolved.
Thus, in Sec.~4.4 we go up to $O(\eps^4)$, 
and derive a higher order averaged equation,
which is capable of describing 
the drift along the incoherent manifold.
Now, $R_2$ converges toward 0 or 1.
This predicts the last remaining multiplier when $N=4$.
In Sec.~4.5 we summarize the results.
Since the calculations are standard, details are shown in
the Appendices.

\subsection{Perturbation expansion}

Chernikov and Schmidt \cite{Che&Sch95} 
recently used the first-order averaging method
to study the governing equations Eq.(\ref{goveqn}) with $\beta>0$,
and obtained the condition for the junction phases
to synchronize to the in-phase state.
It is not clear how to extend their method to higher order.
Here, we employ a multiple time-scale analysis that
allows us to go up to higher order, and
extracts more information on the 
asymptotic dynamics and the phase space structure.

We first rescale the time from $t$ to $\tau$ as
\begin{equation}
  t = \tau / I_b ,
\label{eq4.1.1}
\end{equation}
and rewrite the governing equations Eq.(\ref{goveqn}) as
\begin{equation}
  \left\{ \begin{array}{l}
    \begin{displaystyle}
      m {\phi_j}'' + {\phi_j}' + \eps \sin \phi_j 
        + Q' = 1 ~, ~~ j = 1, \dots, N ,
    \end{displaystyle} \\
    \begin{displaystyle}
      L I_b Q'' + R Q' + Q / (C I_b) = 
        \frac{1}{N} \sum_{k=1}^N {\phi_k}' 
    \end{displaystyle}
  \end{array} \right.
\label{eq4.1.2}
\end{equation}
where $m = I_b \beta$, $\eps = 1/I_b$, and
primes denote derivatives with respect to $\tau$.
Following \cite{Che&Sch95},
we fix $m$, and study the limit $\eps \rightarrow 0+$.
In the pendulum analog of the Josephson junctions, $m$ is a scaled mass
of the pendulum and $\eps$ measures the ratio of the gravity force
to the applied torque.

As usual,
the perturbation expansion will generate secular terms
as we proceed to higher order.
In order to eliminate them,
it turns out that we need to introduce
following three different time scales:
\begin{equation}
  T_0 = \tau ~, ~~ T_2 = \eps^2 \tau ~, ~~ T_4 = \eps^4 \tau ,
\label{eq4.1.4}
\end{equation}
which will be formally treated as independent variables.
Then,
\begin{eqnarray}
  ' & = & \p_0 + \eps^2 \p_2 + \eps^4 \p_4 \nonumber \\
  '' & = & \p^2_0 + 2 \eps^2 \p_0 \p_2 + \eps^4 (\p^2_2 + 2 \p_0 \p_4) 
    + O(\eps^6)
\label{eq4.1.5}
\end{eqnarray}
where $\p_0 = \p / \p T_0$ etc.
We expand the solution in powers of $\eps$:
\begin{equation}
  \phi_j (\tau;\eps) = \sum_{p=0}^\infty \eps^p \phi_{pj} (T_0,T_2,T_4) ~, ~~
  Q (\tau;\eps) = \sum_{p=0}^\infty \eps^p Q_p (T_0,T_2,T_4) ,
\label{eq4.1.6}
\end{equation}
as well as the sinusoidal nonlinearity:
\begin{equation}
  \sin \phi_j = \sum_{p=0}^\infty \eps^p S_{pj}
\label{eq4.1.7}
\end{equation}
where
$S_{0j} = \sin \phi_{0j}$,
$S_{1j} = \phi_{1j} \cos \phi_{0j}$,
$S_{2j} = \phi_{2j} \cos \phi_{0j} - (\phi_{1j}^2/2) \sin \phi_{0j}$,
$S_{3j} = \phi_{3j} \cos \phi_{0j} - (\phi_{1j}^3/6) \cos \phi_{0j}
  - \phi_{1j} \phi_{2j} \sin \phi_{0j}$, and so on.
All $\phi_{pj}$ and $Q_p$ are assumed to be bounded
except $\phi_{0j}$, which increases without bound in time.
Without loss of generality, we impose that the averages
of $\phi_{pj}$, $p>0$, over the fast scale $T_0$ vanish.

\subsection{Fast dynamics}

Substitute Eqs.(\ref{eq4.1.5}--\ref{eq4.1.7}) into (\ref{eq4.1.2}),
and equate like powers of $\eps$.
At the lowest order $O(1)$ we obtain
\begin{equation}
  \left\{ \begin{array}{l}
    \begin{displaystyle}
      m \p^2_0 \phi_{0j} + \p_0 \phi_{0j} + \p_0 Q_0 = 1
        ~, ~~ j = 1, \dots, N ,
    \end{displaystyle} \\
    \begin{displaystyle}
      L I_b \p^2_0 Q_0 + R \p_0 Q_0 + Q_0 / (C I_b) = 
        \frac{1}{N} \sum_{k=1}^N \p_0 \phi_{0k} . 
    \end{displaystyle}
  \end{array} \right.
\label{eq4.1.8}
\end{equation}

This system is linear, inhomogeneous.
Its periodic solutions are easy to find.
\begin{equation}
  \phi^*_{0j} = T_0 + \theta_j ~, ~~ Q^*_0 = C I_b
  \label{eq4.1.9}
\end{equation}
where $\theta_j$ are arbitrary ``constant'' phases
which may, in fact, depend on $T_2$ and $T_4$.
When all $\theta_j$ are identical (or equally spread apart),
the solution is the in-phase (or splay) state.
The (residual) phases $\theta_j$ are same as the phases
obtained by the time-delayed method of Sec.~2.

As shown in Appendix A, it is easy to see that any initial condition
converges to a solution of the form (\ref{eq4.1.9}),
according to the flow (\ref{eq4.1.8}).
The convergence is exponentially fast in the time scale of $T_0$,
and the rate predicts some Floquet exponents.
In particular, the periodic solutions (\ref{eq4.1.9}) must
have (regardless of the choice of $\theta_j$) 
real negative Floquet exponents
\begin{equation}
  \lambda = -2 \pi/m ~~ \mbox{(with multiplicity $N-1$)} .
  \label{eq4.2.101a}
\end{equation}
Also, another set of exponents are obtained by solving
\begin{equation}
  m L I_b \lambda^3 + (m R + L I_b) \lambda^2
  + \left(\frac{m}{C I_b} + R + 1\right) \lambda + \frac{1}{C I_b} = 0 .
  \label{eq4.2.101b}
\end{equation}
This characteristic polynomial produces 1--3 roots
depending on the type of the load,
but their real parts are always negative.

\begin{figure}[tb]
\centerline{\hbox{
\psfig{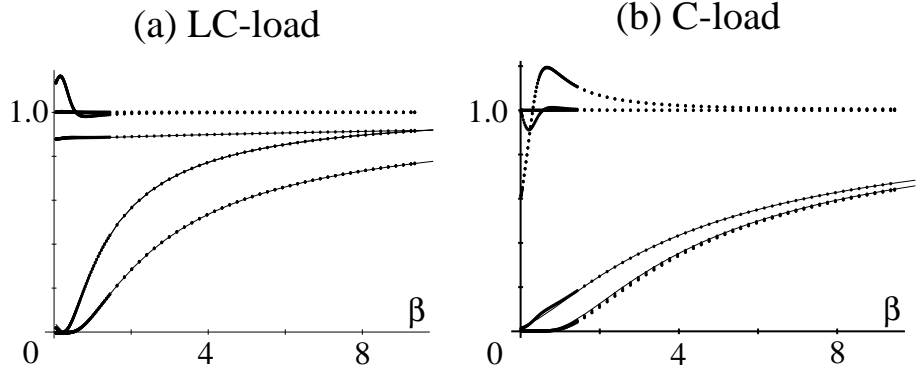}
}}
\caption[comparison1]{
\label{comparison1}
For (a) $LC$-load and (b) $C$-load,
the data points in Fig.~6 whose magnitudes
are well below unity are estimated by the $O(1)$ expansion.
Solid curves are magnitudes corresponding to
Eqs.(\ref{eq4.2.101a},\ref{eq4.2.101b}).
}
\end{figure}
 
Thus, all the exponents except $N$ are estimated
at this order.
In Fig.~7 they are converted to multipliers
$\mu = \exp(2 \pi \lambda)$ (since the period is $2 \pi$),
and shown as solid curves.
(The Floquet multipliers are invariant 
under the scaling of time in (\ref{eq4.1.1}).)
The agreement with the data points located well below the unit magnitude
is excellent even though the formal perturbation parameter $\eps=2/3$ 
is not so small.
These rapidly converging multipliers correspond to the initial
shuffling of oscillators, and after these directions are contracted,
the solution is of the form (\ref{eq4.1.9})
so that all the junctions have the same period.

The remaining $N$ multipliers correspond to perturbation
in the phases $\theta_j$ ($j=1,\dots,N$).
At this order of expansion the junctions are frequency locked,
but there is no preferred phase,
so that any perturbations on $\theta_j$ remain as applied.
Therefore, the $N$ multipliers are all estimated to be unity,
and we need to go up to higher order to resolve
the issue of neutrality.

\subsection{Transverse multipliers}

The arbitrary phases $\theta_j$ are actually dependent
on $T_2$ and $T_4$.
The drift of the phases can be studied
by assuming that the fast dynamics has already contracted,
and $\phi_{0j}$ and $Q_0$ are given by (\ref{eq4.1.9}).
Geometrically, we are only interested in the dynamics
on the center manifold (the invariant $N$-torus)
of the system (\ref{eq4.1.8}).

Since we find no secular terms at the $O(\eps)$ expansion,
we need to consider up to $O(\eps^2)$.
(See Appendix A.)
There, we must impose a solvability condition
to prevent $\phi_{2j}$ from growing unbounded.
The condition is found to be
\begin{equation}
  \partial_2 \theta_j = \Omega_1 + \frac{1}{N} \sum_{k=1}^N
  \left \{ a_1^{~} \cos (\theta_k - \theta_j) 
    + b_1 \sin (\theta_k - \theta_j) \right \}
\label{eq4.3.101}
\end{equation}
where
\begin{equation}
  \Omega_1 = \frac{-1}{2(1+m^2)} ~, ~~
  a_1 = 2 \myRe [c_1] ~, ~~
  b_1 = -2\myIm [c_1] ~, ~~
  c_1 = \frac{1}{4 (1+im) X_1}
\label{eq4.3.102}
\end{equation}
and
\begin{equation}
  X_1 = ( 1+im ) Z_1 + 1 ~,~~
  Z_1 = R + i \left(L I_b - \frac{1}{C I_b} \right) .
\label{eq4.3.103}
\end{equation}
Note that $Z_1$ is the impedance of the load for the frequency $I_b$.
We have not found a physical interpretation for $X_1$ or $c_1$.

Equation (\ref{eq4.3.101}) describes the slow drift of the phases.
It was derived by \cite{Che&Sch95} using the averaging method.
It gives the $O(\eps^2)$ contribution to the normal form for
the dynamics on the center manifold, which is
the invariant $N$-torus. 
We call this normal form the {\it averaged equation}, 
since at each order we remove the next harmonic of the phase modulation
superimposed on the fundamental oscillation.
See \cite{Ash&Swi92}. 
Thus, the second order averaged equation is
\begin{equation}
\label{order2ave}
\dot{\theta_j} = 1 + \eps^2 \left(
\Omega_1 + \frac{1}{N} \sum_{k=1}^N
  \left\{ a_1 \cos (\theta_k - \theta_j)
    + b_1 \sin (\theta_k - \theta_j) \right\}
\right) + O(\eps^4)
\end{equation}
where the dot represents $d/d\tau$.
This has the same form as the averaged equation obtained in
the overdamped limit $\beta \rightarrow 0$
of the governing equation (\ref{goveqn}) \cite{Jai&al84,Wie&Swi95}.

Strictly, the second order equation for $\theta_j$ would look like
Eq.~(\ref{order2ave}), except without the ``1'', which represents a constant
frequency.
The equation for 
$\phi_{0j} = T_0 + \theta_j$ has exactly the form of Eq.~(\ref{order2ave}).
For simplicity, and agreement with the notation in \cite{Ash&Swi92},
we have replaced $\phi_{0j}$ with $\theta_j$ in Eq.~(\ref{order2ave}).

Equation (\ref{order2ave}) is fully nonlinear, but is solvable
by use of a coordinate transformation \cite{Wat&Str94}.
It is proven that
the sign of the parameter $b_1$ determines the attracting set;
when $b_1>0$ the phases $\theta_j$ eventually converges to the in-phase state
while for $b_1<0$ the phases tend to spread so that they converge
to one of the incoherent states.
The first moment $R_1$ defined in Eq.(\ref{defordpar}),
monotonically converges either to $R_1=1$ (in-phase) 
or $R_1=0$ (incoherent), depending on $b_1$.
The critical case $b_1=0$ has been termed by \cite{Che&Sch95} 
the ``synchronization condition''.
In our notation it reduces to
\begin{equation}
  ( 1 - m^2 ) \left ( L I_b - \frac{1}{C I_b} \right )
  + m (1+2R) = 0 .
\label{synccondition}
\end{equation}

Floquet multipliers of the system (\ref{order2ave})
can be evaluated easily
around both the in-phase and the incoherent states.
(See~\cite{Ash&Swi92} and Appendix B here.)
The in-phase state has the unit multiplier, along with
a real Floquet multiplier with multiplicity $N-1$:
\begin{equation}
\label{FmultIn}
\mu_{\mbox{\scriptsize in-phase}} = 
\exp \left( \frac{-\eps^2 2 \pi b_1}{1 + \eps^2 (\Omega_1 + a_1)} \right)
\end{equation}
Assuming $N \geq 3$, the
splay state ($\theta_j$ evenly distributed) in equation~(\ref{order2ave})
has a unit multipier with multiplicity $N-2$,
along with a complex conjugate pair of Floquet multipliers
with magnitude
\begin{equation}
  | \mu_{\pm 1} | = \exp \left (
\frac{\eps^2 \pi b_1}{1+\eps^2 \Omega_1} \right) .
\label{Fmult1}
\end{equation}
These approximate the transverse multipliers
we computed numerically in Sec.~3.
Fig.~8 shows a comparison of Eq.(\ref{Fmult1}) (solid curves
labeled as ``T'')
with the numerical data from Fig.~6.
There is a systematic deviation in the magnitude,
probably due to a large value of $\eps$.
Nonetheless, we see that the formula (\ref{Fmult1}) 
provides a good estimate of the transverse multipliers.
In particular, the transition value of $\beta$
when $| \mu |$ crosses unity is estimated accurately.

\begin{figure}[tb]
\centerline{\hbox{
\psfig{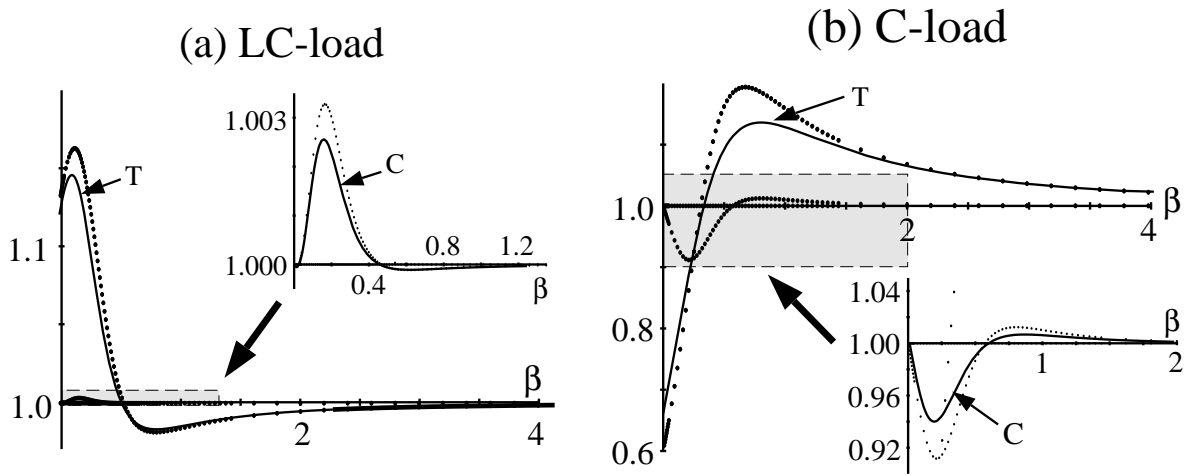}
}}
\caption[comparison2]{
\label{comparison2}
Comparison of the data points in Fig.~6
with the perturbation analysis.
The curves labeled as ``T'' are
the estimate (\ref{Fmult1}) of the  
magnitude of the transverse multipliers.
The curves labeled as ``C'' in the insets are
the formula (\ref{Fmult2}) for the
critical multiplier.
}
\end{figure}
 
When $b_1>0$, the analysis concludes that the in-phase state
is the attractor, and higher order corrections will
provide little information.
However,
when $b_1<0$,
the truncation in equation~(\ref{order2ave}) is too severe.
The splay solution is stable in the transverse direction, but there are
$N-2$ unit multipliers whereas there is generically a single unit multiplier.
The truncated system has an attracting $(N-2)$-dimensional torus,
the incoherent manifold,
foliated with periodic orbits.
Higher order terms break this foliation and lead to generic dynamics.

\subsection{The critical multiplier}

To eliminate the unit multiplers caused by the truncation of
equation~(\ref{order2ave}), we must go to the next order.
By imposing a new solvability condition, we eliminate the secular terms 
which arise at the $O(\eps^4)$ expansion
(since there are no such terms at $O(\eps^3)$).
The calculation is shown in Appendix A.
We find the solvability condition to be:
\begin{equation}
  \p_4 \theta_j = \Omega_2 + \frac{1}{N} \sum_{k=1}^N
  \left\{ a_2 \cos 2 (\theta_k - \theta_j)
    + b_2 \sin 2 (\theta_k - \theta_j) \right\} .
\label{eq4.4.1}
\end{equation}
Here,
\begin{equation}
  \Omega_2 = - \frac{15 m^4 + 10 m^2 + 1}{8 (1+m^2)^3 (1+4 m^2)}
\label{eq4.4.2}
\end{equation}
and $a_2 = 2 \myRe [c_2]$, $b_2 = -2 \myIm [c_2]$ where
\begin{equation}
  c_2 = \frac{i m}{16 X_2 (1-i m) (1+i m)^2 (1+2 i m)}
\label{eq4.4.3}
\end{equation}
and
\begin{equation}
  X_2 = ( 1+2 i m ) Z_2 + 1 ~,~~
  Z_2 = R + i \left(2 L I_b - \frac{1}{2 C I_b} \right) .
\label{eq4.4.4}
\end{equation}
Note that $Z_2$ is the impedance of the load at frequency $2I_b$.

The drift occurs at the time scale of $O(\eps^4)$, and the fourth
order averaged equation is
\begin{equation}
\label{order4ave}
\begin{array}{l}
\begin{displaystyle}
\dot{\theta_j} = 1 +
\eps^2 \left( \Omega_1 + \frac{1}{N} \sum_{k=1}^N
  \left\{ a_1 \cos (\theta_k - \theta_j)
    + b_1 \sin (\theta_k - \theta_j) \right\} \right) 
\end{displaystyle} \\
\begin{displaystyle}
~~~~~~~~~~ + \eps^4 \left( \Omega_2 + \frac{1}{N} \sum_{k=1}^N
  \left\{ a_2 \cos 2(\theta_k - \theta_j)
    + b_2 \sin 2(\theta_k - \theta_j) \right\} \right)  
\end{displaystyle} \\
~~~~~~~~~~ + O(R_1 \eps^4) + O(\eps^6)
\end{array}
\end{equation}
As described in Appendix B, the order $\eps^4$ terms we calculated contribute
to the stability of the splay state.
If $N > 4$ there is a complex conjugate pair of Floquet exponents proportional
to $c_2$ and $\overline{c_2}$.
For $N=4$, the single critical Floquet multiplier is
\begin{equation}
   \mu_{2}  = \exp \left(
\frac{\eps^4 4 \pi b_2}{ 1 + \eps^2 \Omega_1 + \eps^4 \Omega_2} \right) .
\label{Fmult2}
\end{equation}
In the insets of Fig.~8 the predicted critical multiplier is plotted for
both types of the load (curves labeled as ``C''),
and comparison is made with the numerical data from 
the insets of Fig.~6.
Again, we observe a systematic deviation,
but dependence of the multiplier on $\beta$ is 
qualitatively well predicted.
The critical $\beta$ where the splay state
changes its stability 
is estimated accurately.

\subsection{Global dynamics and attractors}

The fourth order perturbation expansion
allows us to predict the attractor for the Josephson
junction system with $N=4$ oscillators.
The incoherent manifold (or incoherent bar)
defined by $R_1=0$ is dynamically
invariant when $N=4$ (see Appendix B),
and we can discuss the drift on the incoherent manifold.
For $N\geq 5$, the set $R_1 = 0$ is not dynamically invariant.

We have concentrated on the Floquet multipliers of the splay state,
but the stability of the in-phase and the 2+2 incoherent states
can also be calculated.
We find that, for $N = 4$ and sufficiently small $\eps$:
\begin{equation}
\label{table}
\begin{array}{l}
\mbox{If}~b_1 > 0 \mbox{~then the in-phase solution is asymptotically stable}\\
\mbox{If}~b_1 < 0 \mbox{~and~} b_2<0 
\mbox{~ then the splay solution is asymptotically stable}\\
\mbox{If}~b_1 < 0 \mbox{~and~} b_2>0 
\mbox{~ then the 2+2 i  solution is asymptotically stable}
\end{array}
\end{equation}

This table does not show the bistability or bi-instability of states, 
which depends
on corrections to the Floquet exponents at higher order in $\eps$.
For example, when the order $\eps^4$ terms in the
averaged equations are truncated,
the Floquet exponents of the splay and
2+2 incoherent states are the same, and these have opposite sign of the
Floquet exponents of the in-phase solution.  All of these exponents are
proportional to $b_1$.  However, when order $\eps^4$ terms are included
in the averaged equations, the exponents for
each of the three solutions gets a different correction.

Similarly, when the order $\eps^6$ terms in the averaged equations
are truncated, the drift on the incoherent manifold
is toward the splay solution ($R_2 \rightarrow 0$) if $b_2 < 0$
and the drift is toward the 2+2 incoherent solution ($R_2 \rightarrow 1$)
if $b_2 > 0$.
However, when order $\eps^6$ terms are included, the exchange of stability
is through a generic incoherent state that is born and dies at separate
pitchfork bifurcations of the splay and 2+2 incoherent states.

To summarize, the asymptotic analysis shows that
$R_1$ and $R_2$ monotonically changes
in separate time scales  in the limit $\eps \rightarrow 0$.
For $N=4$ this determines the attracting periodic
solution uniquely for a given set of parameters;
the possible ones are the in-phase, the splay,
or the 2+2 incoherent state.
Fig. 9 shows the attractor we predict in the
$(I_b,\beta)$ plane, using two different $LC$ loads.

\begin{figure}[tb]
\centerline{\hbox{
\psfig{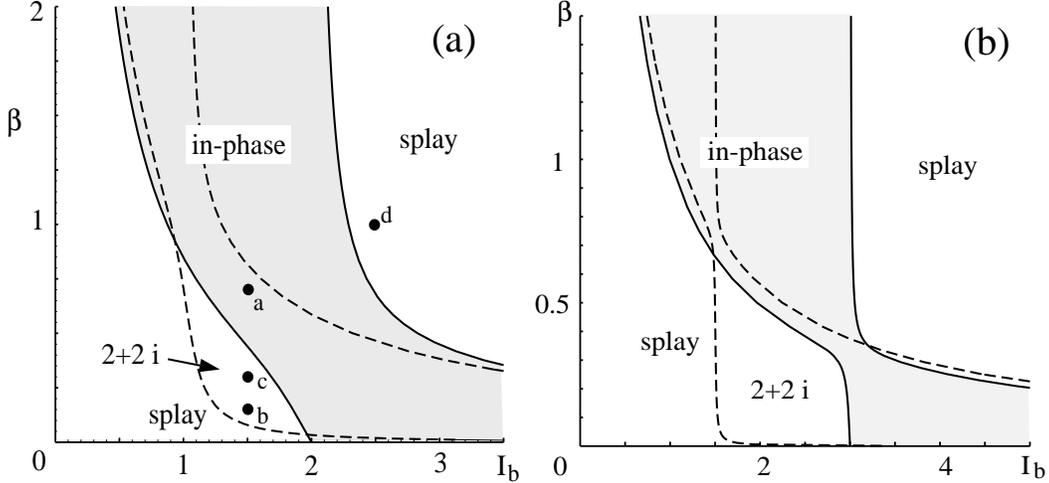}
}}
\caption[parplane]{
\label{parplane}
The asymptotically stable state
predicted for $N=4$.
The perturbation theory assumes $\eps = 1/I_b \ll 1$.
In Fig.~(a), the
same load as in Fig.~2 ($L=1$, $R=0$, $C=1/4$) is used.
Fig.~(b) has $L=100/9$, $R=0$, and $C=1/100$.
The dots in Fig.~(a) marked as ``a'' through ``d'' correspond to
the parameter values used in Fig.~2, respectively.
The solid curve is where $b_1=0$,
and the in-phase state is stable in the shaded region.
Outside the shaded region, the incoherent manifold is attracting.
There is a slow drift on the manifold
whose direction switches on the dashed curve $b_2=0$.
The splay solution is asymptotically stable
in the upper-right region, but the drift is extremely slow.
The point ``b'' where we observed a stable splay state in Fig.~2
is located in the region for the 2+2 incoherent (``2+2 i'') state;
we attribute this to the rather large value of $\eps$.
}
\end{figure}
 
The synchronization condition $b_1 = 0$
determines when the in-phase state is stable.
By substituting $m=I_b \beta$ into (\ref{synccondition}),
we obtain a quadratic equation for $I_b^2$:
\begin{equation}
(L C I_b^2 -1)(\beta^2 I_b^2-1) = \beta C (1 + 2R)
\label{1stboundary}
\end{equation}
For $L>0$, there are (at most) two positive critical $I_b$ values 
for each $\beta$.
The boundary is shown as solid curves, and the region where
the in-phase solution is stable is shaded.
When $\beta = 0$, $b_1 <0$ for $I_b < (LC)^{-1/2}$ and $b_1 > 0$ for
$I_b > (LC)^{-1/2}$.  This agrees with \cite{Wie&Swi95}, and allows us
to determine the sign of $b_1$ in each region.
When $C$ is small, the two factors on the left-hand side of
Eq.~(\ref{1stboundary}) determine the asymptotic boundaries
seen in Fig.~9(b).

The dashed curve, where $b_2 = 0$, concerns the direction of
flow in the incoherent bar.
The condition also becomes a quadratic equation for $I_b^2$:
\begin{equation}
(4 L C I_b^2 -1)(4 \beta^2 I_b^2 - 5) = C[4 \beta(1+4R)-2(1+R)/\beta]
\label{2ndboundary}
\end{equation}
Furthermore, $b_2 = 0$ when $\beta = 0$.
Sign of $b_2$ can be determined from the fact that
$b_2 < 0$ for $\beta$ sufficiently small and positive. 
When $C$ is small,
the two factors on the left-hand side of
Eq.~(\ref{2ndboundary}) determine the asymptotes to $b_2 = 0$
seen in Fig.~9(b).

A plot similar to Fig.~9 can be made for any load (not just an $LRC$ load) by
using the impedance of the load at frequency $I_b$ and $2 I_b$ 
for $Z_1$ and $Z_2$ in equations (\ref{eq4.3.103}) and (\ref{eq4.4.4}).

The bistability we observed numerically in Sec.~2
is not computed by our analysis.
As described after Eq.~(\ref{table}), higher order corrections split the
curve solid curve ($b_1 = 0$) into three curves, and the dashed curve
($b_2=0$) splits in two.

In Fig.~2(d) we used $I_b=2.5$ and $\beta=1$
and observed that the system apparently converged 
to a generic incoherent state.
The splay is asymptotically stable here, but the drift
toward it is too slow to observe.
Farther into the upper-right ``splay'' region of Fig.~9(a),
the coupling between the oscillators is so weak that 
even the dynamics transverse to the incoherent manifold becomes
too slow to observe.  The oscillators ``fall'' into a phase determined
by the initial conditions, and then $R_1$ and $R_2$ appear to remain
constant. 

The lower-left region where the in-phase is unstable
is the most dynamically interesting, since
this is where the moments $R_1$ and $R_2$ evolve quickly.
In the region near $I_b = 1$
there are period doubling bifurcations and bifurcations
to fixed-points and semirotors \cite{Aro&al95} which are not shown in Fig. 9.
The asymptotically stable splay solutions and 2+2 incoherent solutions exist
in the regions with $I_b > 1$.

In Sec.~2.2 we describe the numerical observations along the
line $I_b = 1.5$.  There is qualitative agreement, but the
transition between splay and 2+2 incoherent occurs at 
$\beta \approx 0.2$, higher than in Fig.~9(a).
This discrepancy is not surprising, considering that
$I_b = 1.5$ is not large.
(On the other hand,
recall that the $\beta$ value where $b_2 =0$ is predicted quite accurately
in Fig.~8, where $I_b = 2.5$ and $I_b = 1.5$ with different loads.)
The solid curve in Fig.~9(a) crosses $I_b = 1.5$
within the observed range of bistability
between 2+2 incoherent and in-phase ($0.40 \leq \beta \leq 0.61$).

We have compared our prediction with the numerical
computations of Hadley {\it et al.} \cite{Had&al88}
for the stability of the
in phase state.  The agreement is pretty good,
especially for their Fig.~2(e), where $L=C=1/2$, and $R=0$.
Our figure for this load looks similar to our Fig.~9(a),
since the resonant frequency $(LC)^{-1/2} = 2$ is the same for both loads.
The stability boundary of the in-phase that starts at $I_b = 2$
and arches to the left of $I_b = 1$ agrees surprisingly well with
\cite{Had&al88}.
(This is why we show the region with $I_b < 1$
in Fig.~9.)  For the other $LRC$ loads considered in \cite{Had&al88},
our agreement was good for $I_b$ larger than $1$ or $2$,
but not good for smaller $I_b$.


\section{Summary and discussion}

Globally coupled oscillator systems arise not only in the Josephson
junction context but also in lasers \cite{Ots91,Sil&al92,Rap93},
the complex Ginzburg-Landau equation \cite{Hak&Rap92,Nak&Kur94}, 
and coupled maps \cite{Kan91}.
A ubiquitous feature among these systems are the apparent neutral
stability of the splay state over a wide range of parameters.
In a model of a laser array \cite{Rap93} it is trivial to see that
a foliated incoherent manifold exists, and 
hence that the splay solution is neutrally stable.
Series Josephson arrays with $\beta = 0$ have a similar structure,
but it is more difficult to prove \cite{Wat&Str94}.
On the other hand, for series Josephson arrays with $\beta > 0$
we have shown that the incoherent manifold is not foliated by
incoherent periodic solutions, and that splay solution is not neutrally
stable. 

We have also explained why
previous numerical studies {\em appear} to indicate
that the splay solution is neutrally stable even when $\beta$ becomes finite:
The approximate neutral stability is seen when $\beta$ and/or $I_b$ are
large.
When $I_b$ and $\beta$ are moderate, we can easily find
asymptotically stable splay solutions for an $LC$ or $C$-load.
We have visualized the global dynamics and the drift
on the incoherent manifold (bar) when $N=4$.
Our multiple-scale analysis described this process quantitatively.

Recall that $\eps = 1/I_b$ and $m = I_b \beta$, and our perturbation
expansion assumed $m$ was fixed while $\eps \rightarrow 0$.
This is the natural expansion given the form of the equations,
but a different scaling of the equations shows that
another weak coupling limit is $\beta \rightarrow \infty$
with $I_b$ fixed.
On the other hand,
$\beta \rightarrow 0$ with $I_b$ fixed is {\em not} a weak coupling
limit, because $\Omega_1$ and $c_1$ 
in Eq.(17) have a nonzero limit as $m \rightarrow 0$.

The previous discussion of weak coupling limits makes no assumptions
about the load.  Another way to get weak coupling is to make the load
impedance large.  For example, $R \rightarrow \infty$ is a weak coupling
limit, as exploited in \cite{Wie&Swi95} using a different scaling of the
Josephson junction equations.

The approximate degeneracy which occurs in series Josephson junction array
is due to the simplicity of the averaged equations when {\em truncated}
to order $\eps^2$.
For $N$ oscillators, we must go to order $N$ (if $N$ is even) or $N-1$
(if $N$ is odd) to remove the degeneracy of the truncated averaged equations
(see Appendix B).
For $N=4$ junctions,
the $O(\eps^2)$ equation (\ref{eq4.3.101}) was not sufficient,
but adding the $O(\eps^4)$ corrections in equation~(\ref{eq4.4.1})
remove the degeneracy.

Apart from the theoretical interest
on whether the neutrality is exact or not,
our analysis re-confirms the difficulty of using the splay state
for practical purposes.
Firstly, the parameters must be chosen so that the state is attracting.
For $N=4$ the two conditions $b_1<0$ and $b_2<0$ are required;
for a larger $N$, there will be more conditions to satisfy,
which would make it necessary to tune the parameters precisely.
Secondly, even when this is possible,
the convergence would be very weak.
For $N=4$, the drift of $R_2$ is slow unless the capacitive
load is used and the parameter values are chosen carefully.
For a larger $N$, the drift of the higher phase moments $R_n$ occurs
at a rate proportional to $\eps^{2n}$,
and the splay state is practically neutral in many directions when 
$I_b = 1/\eps$ is large.

It is unknown what happens for large $N$ when $I_b$ is near $1$.
For example, if the parameters are chosen as in Fig.~2(b)
but $N$ is large, is the splay still stable? 
This depends on the signs of $b_n$, for $n$ up to floor[$N/2$].
Based on how $b_1$ and $b_2$ depend on the impedance of the load
at frequency $I_b$ and $2 I_b$, respectively,
we expect that $b_n$ is positive but near $0$ for large $n$,
since the impedance of the load at frequency $n I_b$ is large and inductive.
This will cause the splay to be unstable, with approximate neutral stability.
On the other hand, it is possible that a {\em capacitive} load
has a robustly stable splay solution for any number $N$ of Josephson junctions.

In order to get a robustly stable splay solution for practical applications,
$\beta$ should be small but nonzero, and $I_b$ slightly larger
than one.  Furthermore, a $C$-load should be used (in any case
$L$ should be a few times smaller than $1/C$,
although the coupling will be weak if $1/C$ is too large).
In a different approach to make the splay state more stable,
recent work \cite{Dar&al95,Har&Gar96} describes modifications of
the series array circuit.


\section*{Acknowledgments}

The authors thank Steve Strogatz and Kurt Wiesenfeld for
helpful comments.
JWS acknowledges support by an Organized Research Grant from NAU.
SW was supported in part by NSF grants DMS-9057433 and DMS-9111497
through Steve Strogatz,
and by the A.P.\ Sloan dissertation fellowship.

\appendix

\section{Perturbation expansion}

In this Appendix we show the detail of the calculation in Sec.~4.

\subsection{The fast dynamics}

We start from the setup described in Sec.~4.1.
The $O(1)$ equations are Eq.(\ref{eq4.1.8}).
The stability of the periodic solutions (\ref{eq4.1.9})
can be studied by plugging
\begin{equation}
  \phi_{0j} = \phi^*_{0j} + u_j ~, ~~ Q_0 = Q^*_0 + q ,
\label{eq4.1.10}
\end{equation}
into (\ref{eq4.1.8}) and writing an equivalent linear homogeneous system
for $u_j$ and $q$.
Regardless of the choice of $\theta_j$, we obtain
\begin{equation}
  \left\{ \begin{array}{l}
    \begin{displaystyle}
      m \p^2_0 u_j + \p_0 u_j + \p_0 q = 0
        ~, ~~ j = 1, \dots, N ,
    \end{displaystyle} \\
    \begin{displaystyle}
      L I_b \p^2_0 q + R \p_0 q + q / (C I_b) = 
        \frac{1}{N} \sum_{k=1}^N \p_0 u_k 
    \end{displaystyle}
  \end{array} \right. .
\label{eq4.1.11}
\end{equation}

There are as many independent solutions as the system dimension
($2N+2$ for $LRC$-load and $2N$ for $C$-load)
that grow by a scalar factor $\mu = \exp (2 \pi \lambda)$ 
after one period $2 \pi$ in $T_0$.
The growth factor $\mu$ (or $\lambda$) is a Floquet multiplier 
(or exponent).
Since the system is time independent,
$\lambda$ is simply an eigenvalue of (\ref{eq4.1.11})
and is easy to find.

First, there are $N$ solutions corresponding to perturbations
in phases only:
\begin{equation}
  u_j \equiv \delta_{jn} ~, ~~ \p_0 u_j \equiv 0
  ~, ~~ q \equiv \p_0 q \equiv 0
\label{eq4.1.12}
\end{equation}
for $n=1,\dots,N$
(where $\delta_{jn}$ is the Kronecker delta function).
Each $n$ corresponds to one unit multiplier $\mu=1$, $\lambda=0$.

The second group corresponds to the solutions of the form:
\begin{equation}
  u_j = A_j \exp (\lambda T_0) ~, ~~ q \equiv 0 ~, ~~ \lambda=-1/m 
\label{eq4.1.13}
\end{equation}
which satisfies the first equation in Eq.(\ref{eq4.1.11}).
The second equation can be satisfied by imposing
\begin{equation}
  \sum_{k=1}^N A_k = 0 .
\label{eq4.1.14}
\end{equation}
Since there are $N$ parameters $A_j$ with one constraint (\ref{eq4.1.14}),
there are $N-1$ Floquet multipliers $\mu = \exp (-2 \pi/m)$ from this group.

The last group of multipliers come from the solutions of the form:
\begin{equation}
  u_j = A \exp (\lambda T_0) ~, ~~ 
  q = -A (m \lambda + 1) \exp (\lambda T_0)
\label{eq4.1.15}
\end{equation}
where $A$ is non-zero.
Again, the first equation in Eq.(\ref{eq4.1.11}) is satisfied,
and the second equation can be also satisfied by imposing
Eq.(\ref{eq4.2.101b}).
The characteristic polynomial for $\lambda$ has 3, 2, and 1 root(s)
for an $LRC$-, $RC$-, and $C$-loads, respectively.
Using Hurwitz's criterion, it is easy to show that the real parts of the roots
are always negative.

\subsection{The first solvability condition}

At $O(\eps)$, we obtain
\begin{equation}
  \left\{ \begin{array}{l}
    \begin{displaystyle}
      m \p^2_0 \phi_{1j} + \p_0 \phi_{1j} + \p_0 Q_1 = -\sin \phi_{0j}
        ~, ~~ j = 1, \dots, N ,
    \end{displaystyle} \\
    \begin{displaystyle}
      L I_b \p^2_0 Q_1 + R \p_0 Q_1 + Q_1 / (C I_b) = 
        \frac{1}{N} \sum_{k=1}^N \p_0 \phi_{1k} . 
    \end{displaystyle}
  \end{array} \right. .
\label{eqA.2.1}
\end{equation}
The system is linear, with a forcing from $\phi_{0j}$.
The homogeneous part for $\phi_{1j}$ and $Q_1$ is identical to 
that of the $O(1)$ expansion for $\phi_{0j}$ and $Q_0$.
Thus, the homogeneous solutions consist of
exponentially decaying terms with possible constant 
phases to $\phi_{1j}$.
We set the constants to zero so that the average of $\phi_{1j}$
over $T_0$ becomes zero.

At this order no secularity arises.
The solution is
\begin{equation}
  \left\{ \begin{array}{l}
    \begin{displaystyle}
      \phi_{1j} = \left(
        A_{1j} (T_2,T_4) e^{i T_0} + \mbox{c.c.} \right) / 2
        ~, ~~ j = 1, \dots, N ,
    \end{displaystyle} \\
    \begin{displaystyle}
      Q_1 = \left(
        B_1 (T_2,T_4) e^{i T_0} + \mbox{c.c.} \right) /2
    \end{displaystyle}
  \end{array} \right.
\label{eqA.2.2}
\end{equation}
where ``c.c.'' stands for the complex conjugate, and
\begin{equation}
  \left\{ \begin{array}{l}
    \begin{displaystyle}
      A_{1j} = \frac{1}{1+i m} \left(
        e^{i \theta_j} - \frac{1}{X_1 N}
        \sum_{k=1}^N e^{i \theta_k} \right)
        ~, ~~ j = 1, \dots, N ,
    \end{displaystyle} \\
    \begin{displaystyle}
      B_1 = \frac{1}{X_1 N}
        \sum_{k=1}^N e^{i \theta_k}
    \end{displaystyle}
  \end{array} \right.
\label{eqA.2.3}
\end{equation}
with $X_1$ given in (\ref{eq4.3.103}).

At $O(\eps^2)$ we obtain
\begin{equation}
  \left\{ \begin{array}{l}
    \begin{displaystyle}
      m \p^2_0 \phi_{2j} + \p_0 \phi_{2j} + \p_0 Q_2 = 
        -\p_2 \phi_{0j} -\phi_{1j} \cos \phi_{0j}
        ~, ~~ j = 1, \dots, N ,
    \end{displaystyle} \\
    \begin{displaystyle}
      L I_b \p^2_0 Q_2 + R \p_0 Q_2 + Q_2 / (C I_b) = 
        \frac{1}{N} \sum_{k=1}^N (\p_0 \phi_{2k} + \p_2 \phi_{0k})
    \end{displaystyle}
  \end{array} \right.
\label{eqA.2.5}
\end{equation}

The forcing term on the right hand side of the first equation
involves ``dc'' components (constant of $T_0$).
Unless their sum vanishes,
$\phi_{2j}$ becomes unbounded in $T_0$.
Thus, we impose the solvability condition (\ref{eq4.3.101}).

The solution is then
(neglecting the homogeneous part as in the previous order)
\begin{equation}
  \left\{ \begin{array}{l}
    \begin{displaystyle}
      \phi_{2j} = \left(
        A_{2j} (T_2,T_4) e^{2 i T_0} + \mbox{c.c.} \right) / 2
        ~, ~~ j = 1, \dots, N ,
    \end{displaystyle} \\
    \begin{displaystyle}
      Q_2 = Q_{20} (T_2,T_4) + \left(
        B_2 (T_2,T_4) e^{2 i T_0} + \mbox{c.c.} \right) /2
    \end{displaystyle}
  \end{array} \right.
\label{eqA.2.6}
\end{equation}
where
\begin{equation}
  \left\{ \begin{array}{l}
    \begin{displaystyle}
      A_{2j} = \frac{i/4}{1+2 i m} \left(
        A_{1j} e^{i \theta_j} - \frac{1}{X_2 N}
        \sum_{k=1}^N A_{1k} e^{i \theta_k} \right)
        ~, ~~ j = 1, \dots, N ,
    \end{displaystyle} \\
    \begin{displaystyle}
      Q_{20} = -\frac{C I_b}{4 N} \sum_{k=1}^N
      \left( A_{1k} e^{-i \theta_k} + \mbox{c.c.} \right) ,
    \end{displaystyle} \\
    \begin{displaystyle}
      B_2 = \frac{i/4}{X_2 N}
        \sum_{k=1}^N A_{1k} e^{i \theta_k}
    \end{displaystyle}
  \end{array} \right.
\label{eqA.2.7}
\end{equation}
with $X_2$ given in (\ref{eq4.4.4}).

\subsection{The second solvability condition}

We need the higher order expansions only to resolve the dynamics
along the incoherent manifold.
On this manifold we have
\begin{equation}
  R_1 = \sum_{k=1}^N e^{i \theta_k} = 0
\label{eqA.3.1}
\end{equation}
identically.
Thus, we simplify the terms accordingly.
\begin{equation}
  A_{1j} = e^{i \theta_j} / (1+im) ~, ~~ B_1 \equiv 0
  ~, ~~ Q_1 \equiv 0 
\label{eqA.3.2}
\end{equation}
\begin{equation}
  \p_2 \theta_j = -1/2(1+m^2) = \Omega_1 ~, ~~ Q_{20} = C I_b \Omega_1
\label{eqA.3.3}
\end{equation}
\begin{equation}
  A_{2j} = \frac{i/4}{(1+im) (1+2im)} \left(
    e^{2 i \theta_j} - \frac{1}{X_2 N} \sum_k e^{2 i \theta_k}
  \right) 
\label{eqA.3.3a}
\end{equation}
\begin{equation}
  B_2 = \frac{i/4}{(1+im) X_2 N} \sum_k e^{2 i \theta_k}
\label{eqA.3.3b}
\end{equation}

Now, we write down the $O(\eps^3)$ expansion.
\begin{equation}
  \left\{ \begin{array}{l}
    \begin{displaystyle}
      m \p^2_0 \phi_{3j} + \p_0 \phi_{3j} + \p_0 Q_3
    \end{displaystyle} \\
    \begin{displaystyle}
      =  - 2 m \p_0 \p_2 \phi_{1j} - \p_2 \phi_{1j} 
        - \phi_{2j} \cos \phi_{0j} + (\phi_{1j}^2/2) \sin \phi_{0j}
        ~, ~~ j = 1, \dots, N ,
    \end{displaystyle} \\
    \begin{displaystyle}
      L I_b \p^2_0 Q_3 + R \p_0 Q_3 + Q_3 / (C I_b) = 
        \frac{1}{N} \sum_{k=1}^N \p_0 \phi_{3k}
    \end{displaystyle}
  \end{array} \right.
\label{eqA.3.4}
\end{equation}
The forcing terms in the right hand side only involve 
$\exp (\pm i T_0)$ and $\exp (\pm 3 i T_0)$
so that there is no secularity.
The solution is
\begin{equation}
  \left\{ \begin{array}{l}
    \begin{displaystyle}
      \phi_{3j} = \left(
        A_{31j} (T_2,T_4) e^{i T_0} + 
        A_{33j} (T_2,T_4) e^{3 i T_0} + 
          \mbox{c.c.} \right) / 2
        ~, ~~ j = 1, \dots, N ,
    \end{displaystyle} \\
    \begin{displaystyle}
      Q_3 = \left(
        B_{31} (T_2,T_4) e^{i T_0} +
        B_{33} (T_2,T_4) e^{3 i T_0} +
          \mbox{c.c.} \right) /2
    \end{displaystyle}
  \end{array} \right.
\label{eqA.3.5}
\end{equation}
with
\begin{eqnarray}
  A_{31j} e^{-i \theta_j} & = &
  \frac{2+11 i m-11 m^2}{8 (1+i m)^3 (1-im) (1+2im)} 
  \nonumber \\
  & + &
  \frac{1}{8 (1+im)^2 (1+2im) X_2 N} \sum_{k=1}^N
  e^{2 i (\theta_k - \theta_j)}
\label{eqA.3.6}
\end{eqnarray}
and
\begin{equation}
  B_{31} = 0 .
\label{eqA.3.7}
\end{equation}
(We do not need to write down $A_{33j}$ and $B_{33}$.)

Finally, we consider the $O(\eps^4)$ expansion.
Here, terms independent of $T_0$ appear in the equations
for $\phi_{4j}$, and we will set these terms to vanish.
This condition becomes
\begin{equation}
  \p_4 \theta_j = -
  \left( A_{31j} e^{-i \theta_j}/4
       - A_{1j}^2 \overline{A}_{1j} e^{-i \theta_j}/32
       - i \overline{A}_{1j} A_{2j} e^{-i \theta_j}/8 \right)
  + \mbox{c.c.} 
\label{eqA.3.8}
\end{equation}
which reduces to Eq.(\ref{eq4.4.1}).

\section{Averaged equations and Floquet exponents}

This Appendix concerns the structure of the averaged equations, and 
shows how to pick out the terms that determine the stability of the 
splay solution.

We also describe three special types of averaged equations: pairwise 
coupled systems, centroid coupled systems, and ``the'' averaged equation 
which is both pairwise coupled and centroid coupled.  
Contrary to popular belief, none of these 
special systems describe the weak coupling limit of the series
Josephson junction arrays with $\beta > 0$. 
We conjecture that a centroid coupled system is obtained
in such a limit for the array with $\beta =0$.

We have observed that the perturbation expansion described in Section 4 
and Appendix A leads to an averaged equation of the form
\begin{equation}
\label{calcResult}
\begin{array}{l}
\dot{\theta_j} = 1 + \eps^2
\left[ a + (\alpha e^{-i \theta_j} z_1 + \mbox{c.c.}) \right] \\
 ~~~~~~
+ \eps^4 \left[
       b + c |z_1|^2  
      + \left( e^{-i \theta_j} ( \beta z_1 + \gamma z_1 | z_1|^2 )
      + e^{-2i \theta_j}( \mu {z_1}^2 + \nu z_2 ) + \mbox{c.c.} \right)
     \right] \\
 ~~~~~~
 + O(\eps^6)
\end{array}
\end{equation}
where $a$, $b$,  and $c$ are real constants and $\alpha$, $\beta$, etc. are
complex constants.
The permutation invariant $z_m$ are the moments
of the $N$ oscillators placed on the unit circle,
defined in equation (\ref{defordpar}).

We conjecture that if the the calculation were taken to arbitrary order, the
resulting averaged equation would have the form:
\begin{equation}
\label{aveStructure}
  \dot{\theta_j} = 1 +
  F( \eps e^{i \theta_j}, \eps z_1, \eps^2 z_2, \eps^3 z_3, \ldots),
\end{equation}
where the function $F$ in equation (\ref{aveStructure}) is real-valued and 
invariant under the phase shift:
\begin{equation}
\theta_j \mapsto \theta_j + t , ~~~  z_m \mapsto z_m e^{i m t}
\end{equation}
See section 2 of \cite{Ash&Swi92} for a discussion of this phase shift
symmetry, which is analogous to the phase shift symmetry of a
Hopf bifurcation normal form.
Properties of the dynamics that depend on the phase shift symmetry do not
carry over exactly to the unaveraged equations.
Furthermore, $F$ has the property that
each monomial term
has at least one factor of $\eps e^{\pm i \theta_j}$,
possibly in the form $(\eps e^{i \theta_j})(\eps e^{-i \theta_j}) = \eps^2$.
For example, the term $\eps^2 |z_1|^2$ is found on the right-hand side
of Eq.~({\ref{calcResult}).
The lowest order term of this form is $\eps^4 |z_1|^2$, with coefficient $c$.

We cannot prove that the perturbation expansion to any order leads to
a system of the form~(\ref{aveStructure}).
The reason for our conjecture is this:
For the Josephson junction system, the coupling of the oscillators is through
the current $\dot{Q}$, which is permutation invariant and can be written
in terms of the $z_m$.
There must be at least one exponential factor 
($e^{\pm i \theta_j}$) for the coupling back to
oscillator $j$.
The ordering of the terms as powers of $\eps$ follows because each
exponential factor $e^{i \theta_k}$ is tied to a power of $\eps$.
For example, $z_2$ is quadratic in $e^{i\theta_k}$, and it appears
with a factor of $\eps^2$.  We expect this to continue to higher orders
due to the $\sin(\phi_j)$ type nonlinearity of equation (\ref{goveqn}).

The truncation of terms of order $\eps^4$ and higher in (\ref{calcResult})
leads to a system so common that we have nicknamed it 
``the'' averaged equation.  See (\ref{order2ave}) for ``the'' averaged
equation of the Josephson junction system.

The dynamics of ``the'' averaged equation are
well understood, and highly degenerate.
First of all, there is an invariant incoherent manifold foliated with 
periodic orbits.  Thus, the splay solution is neutrally stable to all 
perturbations within the invariant manifold.
Secondly, the stability of the in-phase solution and the normal 
stability of the incoherent manifold are both determined by the sign of 
$b_1$.  If one solutions is stable, the other is unstable: 
there can be no bistability.  As a result, ``the'' averaged equation
does not have a generic bifurcation as $b_1$ passes through 
$0$.

The main goal of this Appendix is to identify 
the terms in (\ref{aveStructure}),
of order $\eps^4$ and higher, 
that make the dynamics generic. In particular we find those terms that 
contribute to the stability of the splay state.  We shall then make some 
comments concerning the incoherent manifold.

\subsection{Floquet exponents of the splay state}

The splay state is characterized by
\begin{equation}
z_m = \left\{
\begin{array}{ll}
1 & \mbox{if} ~ m = \ldots -2N, -N, 0, N, 2N, \ldots \\
0 & \mbox{otherwise}
\end{array}
\right .
\end{equation}
From the 
product rule, we see that terms in Eq.~(\ref{aveStructure})
with more than one factor of 
$z_m$ or $\overline{z_m}$ with $m \neq pN$
do not contribute to its stability.
The terms with a single $z_m$ factor have the form
$e^{-i m \theta_j} z_m$.
The lowest order term with no factors of
$z_m$ with $m \neq pN$ is
$\eps^{2N} e^{-i N \theta_j} z_N$.
(The $|z_N|^2$ term has order $\eps^{2N+2}$, since there must be a factor
of $|\eps e^{i \theta_j}|^2$.)
Hence, the stability of the splay state is determined by the terms
\begin{equation}
  \dot{\theta_j}  =  c_0 + \left(
  \eps^2 c_1 e^{-i \theta_j}z_1 + 
  \eps^4 c_2 e^{-2 i \theta_j}z_2 + \cdots +
  \eps^{2N} c_N e^{-N i \theta_j}z_N + \mbox{c.c.} \right) +  O(\eps^{2N+2})
\label{order2N}
\end{equation}
The terms listed can all be written in as a ``pairwise coupled'' system:
\begin{equation}
  \dot{\theta_j} =
\frac{1}{N} \sum_{k=1}^N G(\theta_k - \theta_j),
\label{pairwiseCoupled}
\end{equation}
where the real-valued, $2\pi$-periodic coupling function is
\begin{equation}
\label{couplingFunction}
G(\theta) = 
  \sum_{n=-N}^N \eps^{2n} c_{n} e^{i n \theta}
  = c_0 + \sum_{n=1}^N \eps^{2n}
 \{a_{n} \cos(n \theta)  + b_{n} \sin(n \theta) \}
\end{equation}
where $c_{-n} = \overline{c_{n}}$, $a_{n} = 2\mbox{Re}(c_{n})$,
and $b_{n} = -2\mbox{Im}(c_{n})$. 
See equation~(\ref{eq4.4.1}), where the
terms of order $\eps^4$ in the pairwise coupled system are given.

We emphasize that the averaged equations are {\em not} pairwise coupled 
in the Josephson junction system. For example, at order $\eps^4$
there are terms like $e^{-2i\theta_j}z_1^2$ and $e^{-i\theta_j} z_1 | z_1|^2$
on the right hand side of~(\ref{calcResult})
that cannot be written as pairwise coupling.
But if we truncate terms of order $\eps^{2N+2}$, then the stability of
the splay state is determined by pairwise coupled terms.
In fact, we will find that the the splay state is typically hyperbolic
if we truncate terms of order $\eps^{N+1}$.

Note that the Fourier coefficients $c_{n}$ in
(\ref{order2N}) are 
even functions of $\eps$.
For example,
$c_1 = \alpha + \eps^2 \beta + O(\eps^4)$, where $\alpha$ and $\beta$ are
defined in Eq.~(\ref{calcResult}).
In the perturbation expansion, we did not compute the $\beta$ correction,
and we use $c_1$ to stand for the constant $\alpha$ in (\ref{calcResult}).

In this Appendix we concentrate on the Floquet exponents
rather than the multipliers. 
Due to the phase shift symmetry of the averaged equations, the Floquet
exponents are simply the eigenvalues of a constant matrix: hence
we call them $\lambda$.
The Floquet multipliers are $\mu = \exp ( \lambda T)$, where 
$T=2\pi/c_0$ is the period of the oscillation.

Ashwin and Swift \cite{Ash&Swi92} have computed the Floquet
exponents of the splay solutions to pairwise systems.
Fortunately, this result can be used to compute the approximate stability
of the splay when
system~(\ref{aveStructure}) is truncated to order $\eps^{2N}$.
The Floquet exponents of the splay state in~(\ref{pairwiseCoupled}) are
\begin{equation}
  \lambda_m = \frac{1}{N}
  \sum_{k=1}^N G'(2 \pi k/N) (e^{i 2 \pi k m/N} - 1) ,
\end{equation}
where $m$ is any integer satisfying $-N/2 < m \le N/2$.
There are exactly $N$ Floquet exponents;
$\lambda_0 = 0$ always, and if $N$ is
even then $\lambda_{N/2}$ is real.  
The rest of the exponents come in
complex conjugate pairs: $\lambda_{-m} = \overline{\lambda_m}$ for 
$0 < m < N/2$.
Thus we only need consider $1 \le m \le \mbox{floor}[N/2]$.
The exponent $\lambda_m$ corresponds to the perturbation of $z_m$ away
from zero.

It is convenient to cast these results in terms of the Fourier
coefficients of the coupling function $G$, since these are the numbers 
that we get from the perturbation expansion.
The Floquet exponents of the splay solution are
\begin{eqnarray}
\lambda_m & = & \frac{1}{N} \sum_{k=1}^N
\sum_{n = - \infty}^\infty i n c_{n} \eps^{2n} e^{i 2 \pi n k/N}
(e^{i 2 \pi m k/N} - 1) \\
& = & -i \left ( \sum_{n \equiv m} n \eps^{2n} c_{-n} +
\sum_{n \equiv 0} n \eps^{2n} c_{n}
\right)
\end{eqnarray}
The sum over $n \equiv m$ means that all $n$ congruent to $m$ modulo $N$
contribute to the sum.

To compute the first nonzero contribution to the Floquet exponents
the epsilon expansion should be carried out to order
$N$ (if $N$ is even) or $N-1$ (if $N$ is odd).
Then truncate so that $c_{n} = 0$ if $n > N/2$,
and the Floquet exponents of the splay solution are
\begin{equation}
\begin{array}{l}
\lambda_m = -i m \eps^{2m}c_{-m}
\mbox{~~if~~} |m| < N/2, \mbox{~~and} \\
\lambda_{N/2} = (N/2) \eps^N b_{N/2}
\mbox{~~if~$N$~is~even.}  
\end{array}
\end{equation}
The stability of the splay is determined by the real part of the exponent,
or equivalently the magnitude of the Floquet multiplier:
\begin{equation}
\begin{array}{l}
| \mu_{\pm m} | =  \exp(\eps^{2m} m \pi b_m/c_0)
\mbox{~~if~~} |m| < N/2, \mbox{~~and} \\
\mu_{N/2} = \exp(\eps^N N \pi b_{N/2} / c_0)
\mbox{~~if~$N$~is~even.}  
\end{array}
\end{equation}
For $N=4$ oscillators, the critical multiplier $\mu_2$
is determined by the single coefficient $b_2 = -2 \mbox{Re}(c_2)$,
so most of the terms obtained in the averaged equation (\ref{aveStructure})
at order $\eps^4$ can be ignored, as we did in Section 4 by setting $R_1 = 0$
(which is the same as setting $z_1 = 0$).
If one were brave enough to try to compute $c_3$ 
(for $N \geq 6$ oscillators), one
should set $z_1 = z_2 = 0$ for the calculation, and so on to higher order.

\subsection{The incoherent manifold and antiphase pairs}

Is the incoherent manifold invariant?  We show that the answer is no
for $N \geq 5$.  But the set of ``antiphase pairs,'' that coincides
with the incoherent manifold when $N=4$, is shown to
be dynamically invariant in the averaged equations~(\ref{aveStructure}).

Recall that the incoherent manifold is the set $z_1 = 0$.
The manifold is dynamically invariant if $\dot{z}_1 |_{z_1 = 0} = 0$. 
The phase shift
symmetry  and the $\eps$ ordering of equation~(\ref{aveStructure}) imply that
$\dot{z}_1$ is a linear combination of invariant functions times the terms:
\begin{equation}
\label{z1dotTerms}
z_1 , ~~ \eps^2 \overline{z_1} z_2 , ~~ \eps^4 \overline{z_1}^2 z_3,
~~ \eps^4 \overline{z_2} z_1^3, ~~ \eps^4 \overline{z_2} z_3 , ~~ \cdots
\end{equation}
We see that the incoherent manifold is {\em not} dynamically invariant
in general due to terms like $\overline{z_2} z_3$, which 
contain no factors of $z_1$.
The $\overline{z_2} z_3$ term can be traced to the
term proportional to $\overline{\nu}$ in equation~(\ref{calcResult}).

There is, however, a dynamically invariant set.
When $N$ is even, define 
\begin{equation}
{\cal P} = \{ (\theta_1, \cdots, \theta_N)
: ~ \theta_j \mbox{~can be permuted so that~}
\theta_{j+N/2} = \theta_j + \pi  
\mbox{~~for~} j = 1, 2, \ldots, N/2 \}
\end{equation}
In other words, $\cal P$ has $N/2$ antiphase pairs.
It is easy to see that $z_m = 0$ if $m$ is odd for any state in $\cal P$,
because
\begin{equation}
e^{i m \theta} + e^{i m (\theta + \pi)} = 0 \mbox{~~if $m$ is odd.}
\end{equation}
System~(\ref{aveStructure}) can be transformed to an infinite system of
ODEs for the moments ${z}_m$.
The phase shift invariance implies that any polynomial term on the right
hand side of $\dot{z}_m$,
where $m$ is odd, has at least one factor of $z_p$, where $p$
is odd.  Hence $\dot{z}_m|_{\cal P} = 0$ for odd $m$,
and the set $\cal P$ is dynamically invariant in the averaged
equation (\ref{aveStructure}).
We expect a corresponding invariant set in the full Josephson junction
equations, but this is not a rigorous proof, since it relies on
the phase shift symmetry.  (See sec.~2.1 of \cite{Ash&Swi92}.)

For $N = 2$ or $N=4$,
the incoherent manifold {\em is} $\cal P$, so it is invariant in
(\ref{aveStructure}).
The incoherent manifold is also invariant when
$N=3$, since it contains only the splay solution.
For $N \geq 5$, the invariant manifold $z_1 = 0$ is not dynamically
invariant in (\ref{aveStructure}), but since it {\em is} invariant
and normally hyperbolic in the truncation to ``the'' averaged equation,
there is a nearby invariant set when $\eps$ is sufficiently small.

\subsection{The averaged equations when $\beta = 0$}

As we have discussed, the Josephson junction equations with 
McCumber number $\beta = 0$ have a rather degenerate structure.  In
particular, the incoherent manifold is invariant.
We conjecture that the averaged equations in this case are
``centroid coupled:''
\begin{equation}
\label{centroid}
\dot{\theta_j} 
  = \sum_{n=-\infty}^\infty \eps^{2n} c_n e^{-i n \theta_j} z_1^n 
\end{equation}
where $c_{-n} = \overline{c_n}$ are different coefficients from those
in the pairwise coupled system~(\ref{pairwiseCoupled}).
We call this a centroid coupled system because $z_1$ is the centroid 
of the $N$ oscillators placed on the unit circle.  None of the higher 
moments, $z_2, z_3$, etc., appear.  For a centroid coupled system the 
incoherent manifold, defined by $z_1=0$, is dynamically invariant and 
foliated by periodic solutions, all with the same period.  There is no 
``drift'' on the incoherent manifold; in other words all the moments 
$z_m$ are constant when $z_1 = 0$, and the Floquet exponents of the 
splay state are all $0$, except for a single pair giving the stability 
normal to the incoherent manifold.

A centroid coupled system can have bistability (or bi-instability)
between the in-phase solution and the incoherent manifold. For example the 
$e^{-2i \theta_j} z_1^2$ term contributes to the Floquet exponents of the
in-phase solution but not the splay solution. Such bistability is observed 
in a Josephson junction series array with $\beta = 0$ \cite{Tsa&Sch92}.  
Note that a pairwise coupled system cannot describe the array
in the limit, since a pairwise coupled system with
a foliated invariant manifold must be ``the'' averaged equation
({\em i.e.} $c_n = 0$ if $|n| > 1$ in equation~(\ref{centroid})),
which has no bistability.

\end{document}